\documentclass[12pt,english]{article}
\usepackage[T1]{fontenc}
\usepackage[latin9]{inputenc}
\usepackage{geometry}
\usepackage{color}
\usepackage{amsmath}
\usepackage{esint}

\makeatletter

\newcommand{\be}{\begin{equation}}
\newcommand{\ee}{\end{equation}}

\newcommand{\noun}[1]{\textsc{#1}}

\newcommand{\lyxaddress}[1]{
\par {\raggedright #1
\vspace{1.4em}
\noindent\par}
}

\makeatother

\usepackage{babel}
\begin{document}

\title{\noun{Scaling and Renormalization}\\
\noun{in two dimensional Quantum Gravity}}

\author{Alessandro Codello$^{\diamondsuit}$ and Giulio D'Odorico$^{\heartsuit}$}

\maketitle

\lyxaddress{\begin{center}
\emph{$^{\diamondsuit}$CP$^{\,3}$--Origins and Danish IAS,}\\
\emph{University of Southern Denmark,}\\
\emph{Campusvej 55, DK-5230 Odense M, Denmark}\\
\emph{}\\
\emph{$^{\heartsuit}$Radboud University Nijmegen,}\\
\emph{Institute for Mathematics, Astrophysics and Particle Physics,}\\
\emph{Heyendaalseweg 135, 6525 AJ Nijmegen, The Netherlands}
\par\end{center}}

\begin{abstract}
We study scaling and renormalization in two dimensional quantum gravity in a covariant framework.
After reviewing the definition of a proper path integral measure, we use scaling arguments
to rederive the KPZ relations, the fractal dimension of the theory and the scaling of the reparametrization--invariant two point function.
Then we compute the scaling exponents entering in these relations by means of the functional RG.
We show that a key ingredient to obtain the correct results already known from Liouville theory is
the use of the exponential parametrization for metric fluctuations.
We also show that with this parametrization we can recover the correct finite part of the effective action
as the $\epsilon\to 0$ limit of quantum gravity in $d=2+\epsilon$.

\end{abstract}

{\noindent \footnotesize Preprint: CP$^3$-Origins-2014-43 DNRF90 and DIAS-2014-43}

\tableofcontents{}

\section{Introduction}

The study of the quantum nature of gravity still lacks a fundamental
understanding. We now know that the first nontrivial universal quantum
effects, that one can compute considering a free theory with weakly
coupled interactions, are only valid up to an energy scale not exceeding
the Planck scale, and that properly investigating gravity at higher energies
either requires the use of nonperturbative methods, or the introduction
of new physics.

The past decades have witnessed the birth of different nonperturbative
methods that can be used to probe physics beyond the weakly coupled
regime. One of these are functional Renormalization Group (FRG) techniques \cite{Berges:2000ew,Reuter:1996cp} realizing Wilson's
approach to the Renormalization Group (RG). Thanks to this new conceptual paradigm, we now
better understand what are the steps needed to solve a theory.
In the Wilsonian RG, the strengths of the coupling constants in a QFT
form a generalized theory space, and their running describes a trajectory
in this space.
If we want a theory that describes physics at all possible energy
scales, then the endpoints of the RG trajectory cannot sit at any
finite scale, and thus have to represent scale invariant theories.
At a scale invariant point the symmetry of the system gets in many
cases enhanced to the full conformal group, which can be thought of
as generated from Poincarè transformation, dilations and spacetime
inversions. This is strictly true for unitary two dimensional QFTs,
and is probably also true in four dimensions under suitable assumptions.

Critical properties are associated with a scale invariant phase of
the system, since any nonzero mass scale would define a finite correlation
length. Thus critical properties are determined by the fixed points
of the RG flow. Two trajectories with different starting points but
ending at the same fixed point will be associated to the same
critical properties. This fact is called Universality. Universal properties
are the only observable ones, and are associated with a fixed point and
its basin of attraction, that is, with a neighbourhood of it in Theory
Space.
Knowing the critical theory that sits at the conformal point, one
can perturb it and study how it evolves in a neighbourhood of that
point. The critical exponents, which dictate the universal, observable
properties of the theory, are encoded in the conformal data around
the fixed point, namely, for a given representation of the Lorentz
group (spin), in the spectrum of scaling dimensions of the fields
of the theory. This spectrum uniquely determines the lowest order
correlators\footnote{This is only partially true, since one also needs the structure constants $C_{ijk}$.
However, the gravitational dressing will only involve the $\Delta$'s.}, from which all higher order correlators can be reconstructed
using bootstrap relations. In the weakly coupled regime of standard
perturbative QFTs, for example, knowledge of the correlators means
that we can in principle compute the S-matrix elements for a given
process, and thus give the observables.
Thus in order to get the physics one doesn't need to solve the full
theory, but just to look at the universal properties. Reconstructing
the CFT data is one way of accessing the universal properties, and
thus have all the physical predictions of the theory at hand. 

Even then, though, when gravity is introduced its backreaction on
matter will in general change the critical properties of the theory,
and then the CFT data as well, in a nontrivial way. The scaling dimensions
will get a gravitational ``dressing''. The nature of this dressing
is a nonperturbative question, and of fundamental importance if we
want to address the fundamental universal properties of gravity.
If one knows the scaling spectrum at one scale, or for instance at
the fixed point, then one can dress in principle any other scale with
RG flows. The problem is then reduced to that of finding the gravitational
dressing at the fixed point (or near it).

We find that light is sometimes shed on
a complicated problem by investigating a lower dimensional instance
of it.
In two dimensions, in particular, the last 40 years have seen
an incredible progress in our understanding of the mathematical structures
involved. We now know that in two dimensions the bootstrap relations
are so powerful that they allow for an exact solution of the equations,
thus finding the complete spectrum of scaling dimensions. This allows
us to completely classify CFTs. This, together with Zamolodchikov's
proof that any unitary two dimensional scale invariant QFT is a CFT,
implies that we have a complete understanding of the fixed points,
and thus the critical phases of (relativistic) matter in low dimensional
systems.
The last piece of information was given by Knizhnik, Polyakov and 
Zamolodchikov (KPZ), who were able to derive
an exact formula for the gravitational dressing of scaling dimensions \cite{KPZ}.
This tells us the total effect that gravity has on the critical phases of matter, 
and can thus be regarded as a solution of quantum gravity in two dimensions.

In this paper we want to give a unified overview on these aspects,
by using the functional RG. We will see that the main physical ideas
behind them stem from very simple principles, and that the FRG allows
for a very simple derivation of many results that follow from these
arguments.

The paper is organized as follows.
Section 2 introduces the quantities of interest in two dimensional quantum gravity, together with the proper definition of path integral measure that we will use, and then re-derives the relation between the anomaly coefficient and the beta function of Newton's constant.
Section 3 lays down the scaling arguments for two dimensional quantum gravity, in particular deriving the KPZ relations.
Section 4 complements the previous one with the dual approach, namely renormalization.
In the first part we review the details needed in a background field computation, and the basics of functional RG.
In the second part we study gravity in $d=2$: we compute the graviton central charge / coefficient anomaly $c_g$, the scaling exponents needed in the scaling relations, and we comment on the relation with Liouville theory.
In the third and final part we study gravity in $d=2+\epsilon$, deriving in a different way the graviton central charge and analyzing the finite part of the effective action.
Finally, Section 5 is devoted to conclusions.

\section{Two dimensional quantum gravity}

Euclidean two dimensional quantum gravity is loosely defined as the
sum over all metrics living in two dimensional manifolds. Since the
latter, in the closed orientable case, can be topologically classified
by the number of holes $h$, the sum over all metrics becomes a
sum of integrals
over the functional space of metrics on a manifold with fixed topology.
The partition function for $2d$ quantum gravity is then:
\begin{equation}
Z(\Lambda,G)=\sum_{h}\int_{h}\mathcal{D}g\, e^{-\Lambda\int\sqrt{g}+\frac{1}{4\pi G}\int\sqrt{g}R}\,,\label{2dQG_0}
\end{equation}
where $G$ is Newton's constant and $\Lambda$ is the cosmological
constant. Using the Gauss--Bonnet relation $\chi(h)=\frac{1}{4\pi}\int\sqrt{g}R$
we can write:
\begin{align}
Z(\Lambda,G) & =\sum_{h}e^{\chi(h)/G}Z_{h}(\Lambda)\nonumber \\
Z_{h}(\Lambda) & =\int_{h}\mathcal{D}g\, e^{-\Lambda\int\sqrt{g}}\,,\label{2dQG_1}
\end{align}
where $\chi(h)=2-2h$ is the Euler characteristic. We omit to write
the subscript $h$ when $h=0$.

The major problem in defining the path integral over geometries is
the construction or definition of the measure $\mathcal{D}g$ to which
now we turn. We will adopt a pragmatic point of view on the problem
and we will not attempt a rigorous construction, which can be found
in \cite{Ambjorn:1997di}, but instead we want to focus on the basic properties
satisfied by the procedure of averaging over metrics with the intent
of obtaining general scaling relations for the physical quantities.
In other words, we are more interested in computing universal quantities
like critical exponents than solving a particular quantum gravity
model.

\subsection{Fixed--area functionals}

It turns out to be very useful to consider the Laplace transform of
the partition function:
\begin{align}
Z_{h}(\Lambda) & =\int_{0}^{\infty}dA\, Z_{h}(A)\, e^{-\Lambda A}\nonumber \\
Z_{h}(A) & =\int_{h}\mathcal{D}g\,\delta\!\left[\int\sqrt{g}-A\right]\,.\label{2dQG_2}
\end{align}
In fact, these relations show that all we need to do to be able to
compute the partition function is the average of the delta function
containing the composite area operator $\int\!\sqrt{g}$. Note that this
observation is valid for an arbitrary bare action.

Together with the partition function, we can also define other transformed,
or ``fixed-area'', quantities. For instance, the expectation value
of a general operator $\mathcal{O}[\Phi;g]$ depending on some matter fields $\Phi$
 can be rewritten as
\begin{equation}
\left\langle \int\sqrt{g}\,\mathcal{O}[\Phi;g]\right\rangle  =  \sum_{h}e^{\chi(h)/G}\left\langle \int\sqrt{g}\,\mathcal{O}[\Phi;g] \right\rangle _{h}
 = \sum_{h}e^{\chi(h)/G}\int_{0}^{\infty}dA\, F_{h}(A)\, e^{-\Lambda A}\,,
\end{equation}
where $F_{h}(A)$ is the transformed one--point function:
\be
F_{h}(A)=\frac{1}{Z(A)}\int_{h}{\cal D}g\,{\cal D}_{g}\Phi\, \delta\!\left[\int\sqrt{g}-A\right]\,\int\sqrt{g}\,\mathcal{O}[\Phi;g]\,.
\ee
Just as for the partition function, the scaling of $F_{h}(A)$ will determine the full quantum
scaling of the expectation value of the operator, and will tell us how its scaling dimension 
is modified by gravity.

These objects all parametrically depend on the fixed area $A$, whose scaling is still the classical one.
However, we can also consider expectation values depending on more interesting geometrical objects,
such as the geodesic distance between two points.
The geodesic distance is defined as
\be
d_{g}(x,x_{0})=\int_{0}^{1}dt\sqrt{g_{\mu\nu}\left(x\left(t\right)\right)\frac{dx^{\mu}\left(t\right)}{dt}\frac{dx^{\nu}\left(t\right)}{dt}}\,,
\ee
where $x^{\mu}$ is a solution of the geodesic equation:
\be
\frac{d^{2}x^{\mu}\left(t\right)}{dt^{2}}+\Gamma_{\rho\sigma}^{\mu}\left(x\left(t\right)\right)\frac{dx^{\rho}\left(t\right)}{dt}\frac{dx^{\sigma}\left(t\right)}{dt}=0\,,
\ee
with $x(0)=x_{0}$ and $x(1)=x$. 
From this we can construct the (geometric) two--point function \cite{Ambjorn:1997di}:
\be
G\!\left(A,\ell\right)=\int{\cal D}g\, \delta\left(\int \sqrt{g}-A\right)\int d\xi\sqrt{g}\int d\xi^{\prime}\sqrt{g^{\prime}}\delta\left(d_{g}(\xi-\xi^{\prime})-\ell\right)
\label{twopoint}
\ee
whose scaling will also be studied in the following.
The subtle point when studying this quantity is that the parameter $\ell$, which classically
scales as a length, in the full quantum regime acquires a nontrivial scaling.
We can understand this if we notice that the definition of the geodesic distance
involves the Christoffel symbols on the manifold. 
These in the quantum regime become composite operators of
the metric, having their own scaling, which will correct the naive
one.

\subsection{Formal properties of $\mathcal{D}g$}

For any quantum field theory in curved space in $d=2$, a key role
is played by the conformal anomaly. This basically says that the standard
functional measure for matter fields $\mathcal{D}_{g}\Phi$ is not
invariant under Weyl rescalings of the metric $g_{\mu\nu}\rightarrow g_{\mu\nu}e^{2\sigma}$
and the field $\Phi\rightarrow\Phi e^{-\Delta\sigma}$:
\begin{equation}
\mathcal{D}_{ge^{2\sigma}}\left(\Phi\, e^{-\Delta\sigma}\right)=\mathcal{D}_{g}\Phi\, e^{\frac{c_{\Phi}}{24\pi}S_{L}[\sigma;g]}\,,\label{M_1}
\end{equation}
where $S_{L}[\sigma;g]$ is the Wess--Zumino or Liouville action (to
be defined in a moment) and $c_{\Phi}$ is the conformal anomaly coefficient,
or central charge, of the (UV) matter field theory.

The Liouville action is defined as the difference between the Polyakov action,
\begin{equation}
S_{P}[g]=-\frac{1}{96\pi}\int d^{2}x\sqrt{g}\, R\frac{1}{\Delta}R\,,\label{M_2}
\end{equation}
evaluated before and after a Weyl rescaling:
\begin{equation}
S_{P}[ge^{2\sigma}]-S_{P}[g]=-\frac{1}{24\pi}S_{L}[\sigma;g]\,.\label{M_3}
\end{equation}
We thus define the Liouville action as follows:
\begin{equation}
S_{L}[\sigma;g]=\int d^{2}x\sqrt{g}\left[\sigma\Delta\sigma+\sigma R\right]\,.\label{M_4}
\end{equation}
The conformal mode can be eliminated by choosing $\sigma(g)=\frac{1}{2\Delta}R$.
Note that (\ref{M_3}) is invariant modulo a topological term under
constant shifts $\sigma\rightarrow\sigma+\omega$:
\begin{equation}
S_{P}[ge^{2\omega}]=S_{P}[g]-\frac{1}{24\pi}S_{L}[\omega;g]=S_{P}[g]-\frac{1}{6}\chi(h)\omega\,.\label{M_5}
\end{equation}
Following \cite{Codello_D'Odorico_Pagani_2014} we formally construct
a Weyl invariant measure by multiplying the standard one by a factor
$e^{c_{\Phi}S_{P}[g]}$:
\begin{equation}
\mathcal{D}_{g}^{Weyl}\Phi=\mathcal{D}_{g}\Phi\, e^{c_{\Phi}S_{P}[g]}\,.\label{M_6}
\end{equation}
This may be achieved by multiplying by $1=e^{c_{\Phi}S_{P}[g]}e^{-c_{\Phi}S_{P}[g]}$;
thus we can write:
\begin{equation}
\mathcal{D}_{g}\Phi=\mathcal{D}_{g}^{Weyl}\Phi\, e^{-c_{\Phi}S_{P}[g]}\,.\label{M_6.1}
\end{equation}
Here we will make the ansatz, or perform the construction, in which
the same applies to the gravitational case
%
:
\begin{equation}
\mathcal{D}_{g}g=\mathcal{D}_{g}^{Weyl}g\, e^{-c_{g}S_{P}[g]}\,,\label{M_7}
\end{equation}
where $c_{g}$ is the (UV) gravitational anomaly coefficient, or central charge, that has to
be determined self--consistently. The introduction of the Weyl invariant
measure amounts to add to the UV action a Polyakov term $c_{g}S_{P}[g]$.
Thus we can work as if the measure is Weyl invariant, not caring
about the conformal anomaly at all, provided we add the Polyakov action
to the bare action with the correct UV central charge $c_{g}$.
Notice that in order to have a well-defined path integral,
we also need a gauge fixing procedure.
We will leave this implicit until Section 4,
where this issue will be discussed in greater detail.

From now on we will use the Weyl invariant measure and we will drop the label {\it Weyl}, unless otherwise specified (as in Section 4.3.3).

\subsection{Relation between $c_{g}$ and $\partial_{t}G_{k}$}

How do we compute $c_{g}$? The standard way \cite{Distler_Kawai_1988,David_1988} is to use Liouville theory (see Section 4.3.3).
%
Another way is to link $c_{g}$ to the beta function of Newton's constant.
Consider the partition function (\ref{2dQG_1}) at $\Lambda=0$ in which we make a rescaling
of the metric $g_{\mu\nu}\rightarrow\lambda^{-2}g_{\text{\ensuremath{\mu\nu}}}$:
\begin{eqnarray}
Z & = & \sum_{h}e^{\frac{\chi(h)}{G}}Z_{h}\nonumber \\
 & = & \sum_{h}e^{\frac{\chi(h)}{G}}\int_{h}\mathcal{D}\left(\lambda^{-2}g\right)\, e^{-c_{g}S_{P}[\lambda^{-2}g]}\nonumber \\
 & = & \sum_{h}e^{\frac{\chi(h)}{G}}\int_{h}\mathcal{D}g\, e^{-c_{g}S_{P}[g]-\frac{c_{g}}{6}\chi(h)\log\lambda}\nonumber \\
 & = & \sum_{h}e^{\left(\frac{1}{G}-\frac{c_{g}}{6}\log\lambda\right)\chi(h)}Z_{h}\,,\label{M_9}
\end{eqnarray}
where in the first step we made a dummy relabeling and in the second
we used the Weyl invariance of the measure and (\ref{M_5}). This
expression is consistent if Newton's constant is scale dependent, $G=G_{k}$,
evaluated at $\lambda^{-1}k$.
This is to be since we have to keep the physical scale $ds^{2}=g_{\mu\nu}dx^{\mu}dx^{\nu}$ invariant:
the rescaling $g_{\mu\nu}\rightarrow\lambda^{-2}g_{\text{\ensuremath{\mu\nu}}}$
means that the coordinates transform as $x\to\lambda x$, which implies that $k\rightarrow\lambda^{-1}k$.
Since the partition function in this case did not change it must be that:
%
\begin{equation}
\frac{1}{G_{\lambda^{-1}k}}-\frac{c_{g}}{6}\log\lambda=\frac{1}{G_{k}}\,.\label{M_10}
\end{equation}
%
%
%
Expanding (\ref{M_10}) leads to
\begin{equation}
\frac{1}{G_{\lambda^{-1}k}}-\frac{c_{g}}{6}\log\lambda=\frac{1}{G_{k}}-\left[\partial_{t}\left(\frac{1}{G_{k}}\right)+\frac{c_{g}}{6}\right]\log\lambda+...\label{M_10.1}
\end{equation}
where $\partial_t=k \partial_k$. This immediately implies the one--loop exact relation:
\begin{equation}
\partial_{t}\left(\frac{1}{G_{k}}\right)=-\frac{c_{g}}{6} \,,\label{M_11}
\end{equation}
or equivalently:
\begin{equation}
\partial_{t}G_{k}=\frac{c_{g}}{6}G_{k}^{2}\,.\label{M_12}
\end{equation}
These relations show that from the computation of Newton's beta function we can thus determine
the central charge \cite{Codello_D'Odorico_Pagani_2014}.

\section{Scaling}

In this Section we will review the scaling arguments as they follow
from our construction. The key one is the scaling of the expectation
value of a matter (composite) operator, which gives the famous KPZ
dressing \cite{KPZ}, telling us how flat scaling dimensions are modified by gravity.
In order to derive this we will have to start by looking at the scaling
of the partition function. 

\subsection{Partition function}

We will start by presenting the scaling argument for the partition
function $Z_{h}(A)$. We have:
\begin{equation}
Z_{h}(A)=\int_{h}\mathcal{D}g\, e^{-c_{g}S_{P}[g]}\,\delta\!\left(I_{0}[g]-A\right)\,,\label{S_1}
\end{equation}
where $I_{0}[g]\equiv\int\sqrt{g}$ is the area composite operator.
We want to enquire the scaling $A\rightarrow\lambda A$ remembering
that the measure is invariant and the Polyakov action satisfies (\ref{M_5}).
The scaling of the area composite operator can be affected by nontrivial
quantum corrections, hence we will leave it general, $I_{0}[\lambda g]=\lambda^{\alpha}I_{0}[g]$.
We then have:
\begin{eqnarray}
Z_{h}(\lambda A) & = & \int_{h}\mathcal{D}g\, e^{-c_{g}S_{P}[g]}\,\delta\!\left(I_{0}[g]-\lambda A\right)\nonumber \\
 & = & \frac{1}{\lambda}\int_{h}\mathcal{D}g\, e^{-c_{g}S_{P}[g]}\,\delta\!\left(\lambda^{-1}I_{0}[g]-A\right)\nonumber \\
 & = & \frac{1}{\lambda}\int_{h}\mathcal{D}\left(\lambda^{1/\alpha}g\right)\, e^{-c_{g}S_{P}[\lambda^{1/\alpha}g]}\,\delta\!\left(\lambda^{-1}I_{0}[\lambda^{1/\alpha}g]-A\right)\nonumber \\
 & = & \frac{1}{\lambda}\int_{h}\mathcal{D}g\, e^{-c_{g}S_{P}[g]+\frac{c_{g}}{6}\chi(h)\frac{1}{2\alpha}\log\lambda}\,\delta\!\left(I_{0}[g]-A\right)\nonumber \\
 & = & \lambda^{\frac{c_{g}}{12\alpha}\chi(h)-1}Z_{h}(A)\,.\label{S_2}
\end{eqnarray}
If we now choose $\lambda=1/A$, we find the following scaling form:
\begin{equation}
Z_{h}(A)=C_{h}A^{\frac{c_{g}}{6\alpha}(1-h)-1}\,,\label{S_3}
\end{equation}
where $C_{h}=Z_{h}(A=1)$. This is the general form of the fixed-area
partition function, which can now be transformed back to give the
full form of the standard $Z(\Lambda,G)$.

To make further contact with the literature, we can define the string
susceptibility $\gamma$ through:
\begin{equation}
Z_{h}(A)=C_{h}A^{\gamma-3}\,,\label{S_4}
\end{equation}
so that we find
\begin{equation}
\gamma=\frac{c_{g}}{6\alpha}(1-h)+2\,.\label{S_5}
\end{equation}
In the following sections we will see that 
$c_{g}=c_{\Phi}-25$,
%
and we will compute the scaling exponent $\alpha$, obtaining the
value:
\begin{equation}
\alpha=\frac{25-c_{\Phi}-\sqrt{(1-c_{\Phi})(25-c_{\Phi})}}{12}\,.\label{S_6}
\end{equation}
Thus we have:
\begin{equation}
\gamma=-\frac{25-c_{\Phi}+\sqrt{(1-c_{\Phi})(25-c_{\Phi})}}{12}(1-h)+2\,.\label{S_7}
\end{equation}
%
%
In absence of matter we find:
\begin{equation}
\gamma=-\frac{5}{2}(1-h)+2\label{S_8.1}
\end{equation}
and on the sphere\footnote{For a complementary approach based on Matrix Models see \cite{Eichhorn:2013isa}}  $\gamma=-\frac{1}{2}$.

The partition function for two dimensional quantum gravity is then found to be:
\begin{equation}
Z_{h}(\Lambda) =C_{h}\int_{0}^{\infty}dA\, e^{-\Lambda A}A^{\gamma-3}\nonumber =\tau_{h}\Lambda^{2-\gamma}\,,\label{2dQG_6}
\end{equation}
where we defined the model dependent constants $\tau_{h}=C_{h}\Gamma(\gamma-2)$
(see \cite{Ambjorn:1997di} for an explicit evaluation of these constants in
dynamically triangulated gravity). After combining with (\ref{2dQG_1})
we finally arrive at:
\begin{equation}
Z(\Lambda,G)=\sum_{h}\tau_{h}\left(e^{\frac{1}{G}}\Lambda^{-\frac{c_{g}}{12\alpha}}\right)^{\chi(h)}\,,\label{2dQG_7}
\end{equation}
showing that Newton's constant contributes as a ``topological term''
$e^{1/G}$ and that the partition function depends only on the variable
$\kappa=e^{\frac{1}{G}}\Lambda^{-\frac{c_{g}}{12\alpha}}$. In the
case of no matter we have, more explicitly, $\kappa=e^{\frac{1}{G}}\Lambda^{\frac{5}{2}}$.


\subsection{KPZ}

We can apply the same logic to the expectation value of a matter operator
$I_{\mathcal{O}}[\Phi;g]=\int\sqrt{g}\,\mathcal{O}[\Phi;g]$, defined through the fixed-area functional $F_h(A)$.
Suppose the scaling of this operator is $I_{\mathcal{O}}[\Phi,\lambda g]=\lambda^{\beta}I_{\mathcal{O}}[\Phi,g]$.
Again, this takes into account all possible quantum corrections to
a composite operator. Then we have:
\begin{eqnarray*}
F_{h}\left(\lambda A\right) & = & \lambda^{-\frac{c_{g}}{12\alpha}\chi(h)+1}\frac{1}{Z(A)}\int_{h}{\cal D}g\,{\cal D}_{g}\Phi\, e^{-c_{g}S_{P}[g]}\,\delta\!\left(I_{0}[g]-\lambda A\right)\,I_{\mathcal{O}}[\Phi;g]\\
 & = & \lambda^{-\frac{c_{g}}{12\alpha}\chi(h)}\frac{1}{Z(A)}\int_{h}{\cal D}g\,{\cal D}_{g}\Phi\, e^{-c_{g}S_{P}[g]}\,\delta\!\left(\lambda^{-1}I_{0}[g]-A\right)\,I_{\mathcal{O}}[\Phi;g]\\
 & = & \lambda^{-\frac{c_{g}}{12\alpha}\chi(h)}\frac{1}{Z(A)}\int_{h}{\cal D}\left(\lambda^{1/\alpha}g\right){\cal D}_{g}\Phi\, e^{-c_{g}S_{P}[\lambda^{1/\alpha}g]}\,\delta\!\left(\lambda^{-1}I_{0}[\lambda^{1/\alpha}g]-A\right)\,I_{\mathcal{O}}[\Phi,\lambda^{1/\alpha}g]\\
 & = & \lambda^{-\frac{c_{g}}{12\alpha}\chi(h)}\lambda^{\beta/\alpha}\frac{1}{Z(A)}\int_{h}{\cal D}g\,{\cal D}_{g}\Phi\, e^{-c_{g}S_{P}[g]+\frac{c_{g}}{6}\chi(h)\frac{1}{2\alpha}\log\lambda}\,\delta\!\left(I_{0}[g]-A\right)\,I_{\mathcal{O}}[\Phi,g]\\
 & = & \lambda^{\beta/\alpha}\frac{1}{Z(A)}\int_{h}{\cal D}g\,{\cal D}_{g}\Phi\,e^{-c_{g}S_{P}[g]}\,\delta\!\left(I_{0}[g]-A\right)\,I_{\mathcal{O}}[\Phi,g]\\
 & = & \lambda^{\beta/\alpha}F_{h}(A)\,.
\end{eqnarray*}
Choosing again $\lambda=1/A$, we find the scaling form
of the expectation value: 
\begin{equation}
F_{h}(A)=A^{\beta/\alpha}F_{h}(1)\,.
\end{equation}
The physical meaning of this scaling is found by noticing that the
gravitational scaling dimension $\Delta$ can be defined from a one--point
function as $F_{h}(A)\propto A^{1-\Delta}$,
while $\beta$ is related to the flat scaling dimension $\Delta_{0}$
of the operator ${\cal O}$, as
we will prove later, by
\begin{equation}
\beta=\frac{25-c_{\Phi}-\sqrt{(1+24\Delta_{0}-c_{\Phi})(25-c_{\Phi})}}{12}\,.
\end{equation}
This means that the scaling dimension $\Delta_{0}$ of an operator
receives a gravitational dressing which changes it into
\be
\Delta=1-\frac{\beta}{\alpha}=\frac{\sqrt{1-c_{\Phi}}-\sqrt{1+24\Delta_{0}-c_{\Phi}}}{\sqrt{1-c_{\Phi}}-\sqrt{25-c_{\Phi}}} \,,
\ee
which is the KPZ formula.
This relation can be recast in the equivalent form (see Section 4.3.2):
\be
\Delta-\Delta_{0}=\frac{6\alpha^{2}}{25-c_{\Phi}}\Delta(\Delta-1)\,,
\ee
also known in the literature as KPZ relation, which shows clearly
that all the effect of gravity is encoded in the scaling $\alpha$
of the area operator.

\subsection{Fractal properties of spacetime}

The previous considerations only required the scaling of a fixed area,
which is dictated by its classical scaling. However, if the partition
function starts to depend on less trivial geometrical quantities such
as the geodesic distance between two points, as in the case of the
two point function (\ref{twopoint}), the scaling of these quantities
can get a nontrivial modification with respect to the classical one,
as we here briefly review \cite{Watabiki:1993fk}.

The effective scaling dimension in a quantum spacetime can be probed
by considering a random walk, or a diffusion process, and studying its
properties. The scaling dimension is related to the return probability,
which in our case can be expressed as:
\be
P(A;s)=\left\langle \frac{1}{A}{\rm Tr}\, e^{-s\Delta}\right\rangle =\frac{1}{Z(A)}\int{\cal D}g\, e^{-c_{g}S_{P}[g]}\,\delta\!\left(I_{0}[g]-A\right)\,\frac{1}{A}{\rm Tr}\, e^{-s\Delta}\,,
\ee
in which $s$ is the diffusion time,
and $K_{s}= e^{-s\Delta}$ is the heat kernel,
which is a solution of the heat equation \cite{Codello_Zanusso_2013}:
\be
\partial_{s}K_{s}(x,x_{0})+\Delta_{x}K_{s}(x,x_{0})=0\,,
\ee
with boundary condition $K_{0}(x,x_{0})=\delta(x-x_{0})/\sqrt{g}$. 
The scaling dimension in the
UV is related to the way in which $P(A;s)$ scales as a function of
$s$ for $s\to0$. We immediately notice that at $s=0$ we have
\be
P(A;0)=1=P(\lambda A;0)\,.
\ee
If we assume that this holds also for small finite $s$, whose scaling
is still unknown, by repeating the same manipulations of the previous
sections we find the following relation:
\begin{eqnarray*}
P(\lambda A;\lambda^{\omega}s) & = & \frac{1}{Z(A)}\int{\cal D}g\, e^{-c_{g}S_{P}[g]}\,\delta\!\left(I_{0}[g]-A\right)\,\frac{1}{A}{\rm Tr}\, e^{-\lambda^{\omega}s\Delta(\lambda^{1/\alpha}g)}\\
 & = & \frac{1}{Z(A)}\int{\cal D}g\, e^{-c_{g}S_{P}[g]}\,\delta\!\left(I_{0}[g]-A\right)\,\frac{1}{A}{\rm Tr}\, e^{-\lambda^{\omega+\tilde{\alpha}/\alpha}s\Delta(g)}\\
 & = & P(A;s)\,,
\end{eqnarray*}
where we used the fact that ${\rm Tr}\,K_s(x,y)=\int \sqrt{g}\,{\rm tr}\,K_s(x,x)$,
so it scales as $I_{0}[g]$, and the Laplacian has its own scaling
$\Delta(\lambda g)=\lambda^{\tilde{\alpha}}\Delta(g)$, with $\tilde{\alpha}$ a new scaling exponent. The only way
to fulfil this condition is that $\omega=-\tilde{\alpha}/\alpha$.
Thus the diffusion time $s$ scales as $A^{-\tilde{\alpha}/\alpha}$.
The average, in the diffusion process, of the squared geodesic distance
from a starting point $x_{0}$ is given by (the subscript $s$ indicates
that this is the diffusion average, not the quantum one):
\be
\left\langle d_{g}^{2}(x,x_{0})\right\rangle _{s}=\int d^{2}x\sqrt{g}\,d_{g}^{2}(x,x_{0})K_{s}(x,x_{0})\,.
\ee
The scaling is determined by the small $s$ behaviour of the average.
Expanding the heat kernel for small $s$, using $d_{g}(x_{0},x_{0})=0$,
we see that $\left\langle d_{g}^{2}(x,x_{0})\right\rangle _{s}$ starts
at linear order in $s$, and thus scales as well as $A^{-\tilde{\alpha}/\alpha}$.
The coefficient $\tilde{\alpha}$ will be determined later like the
other exponents encountered previously. We will find that
\be
\frac{\tilde{\alpha}}{\alpha}=\frac{\sqrt{49-c_{\Phi}}-\sqrt{25-c_{\Phi}}}{\sqrt{1-c_{\Phi}}-\sqrt{25-c_{\Phi}}}\,.
\ee
This relation (or more precisely an equivalent version of it) was found in \cite{Watabiki:1993fk}.
We see that for $c=0$ we have $\tilde{\alpha}/\alpha=-1/2$, which
means that in the full quantum regime the geodesic distance scales
like $A^{1/4}$.

A more direct physical way of seeing this can be the following. We
know that for a random walk on a fractal, the average square displacement
is related to the walking time $T$ by the power--law:
\be
\left\langle r^{2}\right\rangle \propto T^{2/d_{w}}\,,
\ee
$d_{w}$ being the walking dimension. Since the walking time scales
like an area (this is a general property of random walks in two dimensions: their trajectories
have Hausdorff dimension 2), we deduce that the geodesic distance
scales like $d_{g}\sim A^{1/d_{w}}$. Now we can use the known form
of $d_{w}$ \cite{Reuter:2011ah}, which is
\be
d_{w}=4+\textrm{beta--functions}
\ee
to get that, at a fixed point, $d_{w}^{*}=4$, and thus $d_{g}$ scales
like $A^{1/4}$.


\subsection{Two point function}

Finally, knowing the scaling of the geodesic distance, we can reproduce
the previous arguments for the geometric two point function $G\!\left(A,\ell\right)$ defined in (\ref{twopoint}).
Using the scaling we just found for the geodesic length, we find:
\begin{eqnarray*}
G\!\left(\lambda A,\,\lambda^{1/4}\ell\right) & = & \int{\cal D}g\, e^{-c_{g}S_{P}[g]}\,\delta\left(I_{0}\left[g\right]-\lambda A\right)\int d^{2}\xi\sqrt{g}\int d^{2}\xi^{\prime}\sqrt{g^{\prime}}\delta\left(d_{g}(\xi-\xi^{\prime})-\lambda^{1/4}\ell\right)\\
 & = & \lambda^{-\frac{5}{4}}\int{\cal D}g\, e^{-c_{g}S_{P}[g]}\,\delta\left(\lambda^{-1}I_{0}\left[g\right]-A\right)\int d^{2}\xi\sqrt{g}\int d^{2}\xi^{\prime}\sqrt{g^{\prime}}\delta\left(\lambda^{-1/4}d_{g}(\xi-\xi^{\prime})-\ell\right)\\
 & = & \lambda^{-\frac{5}{4}}\int{\cal D}\left(\lambda^{1/\alpha}g\right)\, e^{-c_{g}S_{P}[\lambda^{1/\alpha}g]}\,\delta\left(\lambda^{-1}I_{0}\left[\lambda^{1/\alpha}g\right]-A\right)\times\\
 &  & \lambda^{2}\int d^{2}\xi\sqrt{g}\int d^{2}\xi^{\prime}\sqrt{g^{\prime}}\delta\left(\lambda^{-1/4}d_{\lambda^{1/\alpha}g}(\xi-\xi^{\prime})-\ell\right)\\
 & = & \lambda^{\frac{3}{4}+\frac{c_{g}}{12\alpha}\chi(h)}\int{\cal D}g\, e^{-c_{g}S_{P}[g]}\,\delta\left(I_{0}\left[g\right]-A\right)\int d^{2}\xi\sqrt{g}\int d^{2}\xi^{\prime}\sqrt{g^{\prime}}\delta\left(d_{g}(\xi-\xi^{\prime})-\ell\right)\\
 & = & \lambda^{\frac{3}{4}+\frac{c_{g}}{12\alpha}\chi(h)}G\!\left(A,\ell\right)\,,
\end{eqnarray*}
in which we had to assume that $\lambda^{-1/4}d_{\lambda^{1/\alpha}g}(\xi-\xi^{\prime})=d_{g}(\xi-\xi^{\prime})$,
which is required to have a well defined delta function.
Taking again $\lambda=A^{-1}$ we find:
\be
G\!\left(A,\ell\right)=A^{\frac{3}{4}+\frac{c_{g}}{12\alpha}\chi(h)}f(\ell A^{-1/4})\,,
\ee
with $f(x)\equiv G(1,x)$. The scaling
in $\Lambda$ will then be:
\begin{eqnarray*}
G\!\left(\Lambda,\ell\right) & = & \int_{0}^{\infty}dA\, G\left(A,\ell\right) \, e^{-\Lambda A}\\
 & = & \int_{0}^{\infty}dA\, A^{\frac{3}{4}+\frac{c_{g}}{12\alpha}\chi(h)}f(\ell A^{-1/4}) \, e^{-\Lambda A}\\
 & = & \Lambda^{-\frac{7}{4}-\frac{c_{g}}{12\alpha}\chi(h)}g(\ell\Lambda^{1/4})\,,
\end{eqnarray*}
with
\be
g(x)\equiv\int_{0}^{\infty}dy\, e^{-y}\, y^{3/4+\frac{c_{g}}{12\alpha}\chi(h)}f(x\, y^{-1/4})\,.
\ee
This way we recover the known scaling on the sphere \cite{Ambjorn:1997di}:
\begin{equation}
G\!\left(\Lambda,\ell\right) = \varLambda^{-\frac{7}{4}+\frac{5}{2}(1-h)}g(\ell\Lambda^{1/4}) \underset{(h=0)}{=}  \varLambda^{\frac{3}{4}}g(\ell\Lambda^{1/4})\,.
\label{scalingG}
\end{equation}
The scaling function has been computed in \cite{Ambjorn:1997di} and its detailed
form is $g(x)=\cosh x / \sinh^{3}x$.
We remark that our scaling relation is valid also for $c_{\Phi} \neq 0$.

\section{Renormalization}

In this Section we will consider renormalization in order to compute
the anomaly coefficient $c_{g}$ and the critical exponents $\alpha,\beta,...$
that characterize the scaling laws derived in the previous Section.
In particular we will compute the beta function of Newton's constant
since this leads to the computation of $c_{g}$ via relation (\ref{M_11}).
The critical exponents are instead related to the scaling dimensions
of composite operators, such as $I_{0}[g]=\int\sqrt{g}$.

We will perform our computations in both $d=2$ and $d=2+\epsilon$
in order to enquire various things: which operator drives the flow,
i.e. $S_{P}[g]$ in $d=2$ versus $\int\sqrt{g}R$ in $d=2+\epsilon$;
the connection with Liouville theory; the limit $\epsilon\rightarrow0$;
the role of different parametrizations of the metric fluctuation. 
From now on we will also fix the topology to be spherical, since 
$c_{g}$ and the scaling exponents do not depend on the topology.

\subsection{Background, gauges and ghosts}

We need now to discuss in more detail the construction of the measure.
The standard approach, that we will follow here, is the original Faddeev--Popov
method that allows to factor out the volume of the $Diff$--group via
a gauge--fixing and at the cost of introducing ghost fields, or better
at the cost of introducing an additional functional determinant:
\begin{equation}
\mathcal{D}g \to \mathcal{D}g\,\delta[f]\, Z_{gh}[g]\,,\label{PI_1}
\end{equation}
where $f=0$ is the gauge--fixing condition and $Z_{gh}[g]$
is the Fadeev--Popov determinant. A nice and elegant introduction to gravitational
functional integrals can be found in \cite{Mottola:1995sj}, to which we refer for more details.

To preserve invariance under diffeomorphisms we employ the background
field method where we expand around a background metric $\bar{g}_{\mu\nu}$
and we integrate over the metric fluctuation $h_{\mu\nu}$. Fluctuations
can be parametrized in different ways; here we will discuss two of
them, the linear (or standard) parametrization:
\begin{equation}
g_{\mu\nu}=\bar{g}_{\mu\nu}+h_{\mu\nu}\,,\label{PI_1.01}
\end{equation}
and the exponential parametrization \cite{Kawai_Kitazawa_Ninomiya_1992,Eichhorn:2013xr,Nink_2014}:
\begin{equation}
g_{\mu\nu}=\bar{g}_{\mu\lambda}e^{h_{\;\nu}^{\lambda}}=\bar{g}_{\mu\nu}+h_{\mu\nu}+\frac{1}{2}h_{\mu\lambda}h_{\;\nu}^{\lambda}+...\,.\label{PI_1.02}
\end{equation}
Since the flow equations involve the Hessians, or second variations,
of the effective action, we will keep track of the parametrization
choice by introducing the tensor $H_{\mu\nu}=\xi h_{\mu\lambda}h_{\;\nu}^{\lambda}$,
where $\xi$ is a parameter which can be either zero or one, so that we can write $\delta g_{\mu\nu}=h_{\mu\nu}$
and $\delta^{2}g_{\mu\nu}=H_{\mu\nu}$. Now the functional integration
over the metric $g_{\mu\nu}$ can be replaced by one over the fluctuation
$h_{\mu\nu}$, i.e. $\mathcal{D}g_{\mu\nu}=\mathcal{D}h_{\mu\nu}$
and the Fadeev--Popov operator $\mathcal{M}$, defined by $\det\mathcal{M} = Z_{gh}[g]$, is given by:
\begin{equation}
\mathcal{M}[h;\bar{g}]=\left.\frac{\delta f[h^{\epsilon},\bar{g}]}{\delta\epsilon}\right|_{\epsilon=0}\,,\label{PI_1.03}
\end{equation}
where $h_{\mu\nu}^{\epsilon}=h_{\mu\nu}+\nabla_{\mu}\epsilon_{\nu}+\nabla_{\nu}\epsilon_{\mu}$
represents an infinitesimal coordinate transformation of the tensor $h_{\mu\nu}$
with respect to the full metric $g_{\mu\nu}$. There are now
two possible gauge choices. The conformal gauge (CG)
\begin{equation}
f_{\mu\nu}=h_{\mu\nu}-\frac{1}{2}\bar{g}_{\mu\nu}h\,,\label{PI_1.1}
\end{equation}
that fixes the gauge completely only $d=2$, and the Feynman gauge (FG)
\begin{equation}
f_{\mu}=\bar{\nabla}^{\nu}h_{\mu\nu}-\frac{1}{2}\bar{\nabla}_{\mu}h\,,\label{PI_1.2}
\end{equation}
which can be used in any dimension $d\geq2$. Note that the FG is
the gradient of the CG. As usual the strict gauge--fixing condition
$\delta[f]$ can be relaxed by exponentiation of the delta function,
in this way introducing the gauge--fixing action $S_{gf}[h;\bar{g}]$,
that in the background gauge depends on both the fluctuation and the
background metric. In CG the gauge--fixing action is:
\begin{equation}
S_{gf}[h;\bar{g}]=\frac{1}{2\alpha}\int d^{2}x\sqrt{\bar{g}}\left(h_{\mu\nu}-\frac{1}{2}\bar{g}_{\mu\nu}h\right)\left(h^{\mu\nu}-\frac{1}{2}\bar{g}^{\mu\nu}h\right)\,,\label{PI_1.3}
\end{equation}
while in FG it is:
\begin{equation}
S_{gf}[h;\bar{g}]=\frac{1}{2\alpha}\int d^{d}x\sqrt{\bar{g}}\left(\bar{\nabla}^{\alpha}h_{\alpha\mu}-\frac{1}{2}\bar{\nabla}_{\mu}h\right)\left(\bar{\nabla}^{\beta}h_{\beta}^{\mu}-\frac{1}{2}\bar{\nabla}^{\mu}h\right)\,.\label{PI_1.4}
\end{equation}
In both cases $\alpha$ is the gauge--fixing parameter%
\footnote{The gauge--fixing parameter should not be confused with the scaling
exponent labeled in the same way.%
}.

The Fadeev--Popov operator $\mathcal{M}$ can be computed given the
gauge condition. In CG the variation of the gauge condition leads
to (remember that $h=\bar{g}^{\mu\nu}h_{\mu\nu}$):
\begin{equation}
\delta f_{\mu\nu}=\delta h_{\mu\nu}-\frac{1}{2}\bar{g}_{\mu\nu}\delta h=\nabla_{\mu}\epsilon_{\nu}+\nabla_{\nu}\epsilon_{\mu}-\bar{g}_{\mu\nu}\nabla\cdot\epsilon\equiv(L\epsilon)_{\mu\nu}\,,\label{PI_2}
\end{equation}
which defines the vector to symmetric traceless rank two tensor differential
operator $L_{\mu\nu}^{\alpha}\equiv\delta_{\nu}^{\alpha}\nabla_{\mu}+\delta_{\mu}^{\alpha}\nabla_{\nu}-\bar{g}_{\mu\nu}\nabla^{\alpha}$.
Introducing the adjoint operator,
\begin{equation}
\int\sqrt{\bar{g}}\chi^{\mu\nu}(L\epsilon)_{\mu\nu}=2\int\sqrt{\bar{g}}\chi^{\mu\nu}\nabla_{\mu}\epsilon_{\nu}=-2\int\sqrt{\bar{g}}\,\nabla_{\mu}\chi^{\mu\nu}\epsilon_{\nu}\equiv\int\sqrt{\bar{g}}\,(L^{\dagger}\chi)^{\nu}\epsilon_{\nu}\,,\label{PI_3}
\end{equation}
we find $(L^{\dagger})_{\alpha}^{\mu\nu}=-(\delta_{\alpha}^{\mu}\nabla^{\nu}+\delta_{\alpha}^{\nu}\nabla^{\mu})$.
We can handle the FP determinant more easily using the
fact that $L$ and $L^{\dagger}$ have the same non--zero eigenvalues:
\begin{equation}
\det\mathcal{M}=\textrm{det}'L=(\textrm{det}'L^{\dagger}L)^{\frac{1}{2}}\,,\label{PI_4}
\end{equation}
where we exclude the zero modes from the determinant, which
are actually the zero modes of $L^{\dagger}$. It's easy to reveal
the explicit form of the FP operator when $h_{\mu\nu}=0$:
\begin{eqnarray}
(L^{\dagger})_{\beta}^{\mu\nu}L_{\mu\nu}^{\alpha} & = & -(\delta_{\beta}^{\mu}\bar{\nabla}^{\nu}+\delta_{\beta}^{\nu}\bar{\nabla}^{\mu})(\delta_{\nu}^{\alpha}\bar{\nabla}_{\mu}+\delta_{\mu}^{\alpha}\bar{\nabla}_{\nu}-\bar{g}_{\mu\nu}\bar{\nabla}^{\alpha})\nonumber \\
 & = & 2(\bar{\Delta}\delta_{\beta}^{\alpha}-\bar{R}_{\beta}^{\alpha})\equiv2(\bar{\Delta}_{1})_{\beta}^{\alpha}\,,\label{PI_5}
\end{eqnarray}
where we introduced the spin one Laplacian $\Delta_{1}$. In FG we
have instead:
\begin{equation}
\delta f_{\mu} = 
 \left(\delta_{\mu}^{\alpha}\bar{\nabla}^{\beta}-\frac{1}{2}\bar{g}^{\alpha\beta}\bar{\nabla}_{\mu}\right)\delta h_{\alpha \beta}\nonumber \\
 =  \left(\delta_{\mu}^{\alpha}\bar{\nabla}^{\beta}-\frac{1}{2}\bar{g}^{\alpha\beta}\bar{\nabla}_{\mu}\right)\left(\nabla_{\alpha}\epsilon_{\beta}+\nabla_{\beta}\epsilon_{\alpha}\right)\,,
\end{equation}
%
%
thus:
\begin{equation}
\det\mathcal{M}=\det\left(\bar{\nabla}^{\alpha}g_{\alpha\nu}\nabla_{\mu}+\bar{\nabla}^{\alpha}g_{\mu\nu}\nabla_{\alpha}-\bar{\nabla}_{\mu}g_{\nu\alpha}\nabla^{\alpha}\right)\,.\label{PI_7}
\end{equation}
Note that the differential operator in (\ref{PI_7}) depends non trivially
on $h_{\mu\nu}$ and $g_{\mu\nu}$. It simplifies when $h_{\mu\nu}=0$:
\begin{equation}
\bar{\nabla}^{\alpha}\bar{g}_{\alpha\nu}\bar{\nabla}^{\mu}+\bar{\nabla}^{\alpha}\delta_{\nu}^{\mu}\bar{\nabla}_{\alpha}-\bar{\nabla}^{\mu}\bar{g}_{\nu\alpha}\bar{\nabla}^{\alpha}=\bar{\Delta}\delta_{\nu}^{\mu}-\bar{R}_{\nu}^{\mu}=(\bar{\Delta}_{1})_{\beta}^{\alpha}\,.
\label{PI_8}
\end{equation}
We see that the CG and FG ghost operators are the same when $h_{\mu\nu}=0$,
but CG has a real ghost while FG as complex ghosts, i.e. the determinant
is under square root in the first case. Note also that in CG we need to exclude
zero modes, while in FG we need not (since it is the gradient of the
CG).

A final comment on the CG in $d=2$. In the exponential parametrization,
i.e. $\xi=1$, we have:
\begin{equation}
g_{\mu\nu}=\bar{g}_{\mu\lambda}e^{h_{\:\nu}^{\lambda}}=\bar{g}_{\mu\lambda}e^{\frac{1}{2}\delta_{\nu}^{\lambda}h}=\bar{g}_{\mu\lambda}\delta_{\nu}^{\lambda}e^{\frac{1}{2}h}=\bar{g}_{\mu\nu}e^{\frac{1}{2}h}\equiv\bar{g}_{\mu\nu}e^{2\sigma}\,,\label{CG_2}
\end{equation}
showing that all metrics can be reached from the background metric
via a Weyl transformation with factor:
\begin{equation}
\sigma=\frac{h}{4}\,.\label{CG_3}
\end{equation}
Note that the Liouville action (\ref{M_4}) in terms of $h$ reads:
\begin{eqnarray}
\frac{1}{24\pi}S_{L}[h/4;g] & = & \frac{1}{96\pi}\int d^{2}x\sqrt{g}\left[\frac{1}{4}h\Delta h+hR\right]\,.\label{L_3.1}
\end{eqnarray}
Having set up the background, the gauges and the ghosts, we can now
turn to the discussion of the RG flow equations.

\subsection{Flow equations}

To study the renormalizaion group flow and to compute the beta function
and the critical or scaling exponents we will employ the functional
RG approach based on the exact RG flow equation satisfied by the effective
action\footnote{The effective action, when evaluated on--shell, is related 
to the partition function by $\Gamma_{*}=-\log Z$.} \cite{Reuter:1996cp,Codello:2008vh}.

Using the background field method to preserve gauge invariance along
the flow leads to the following flow equation first derived in  \cite{Reuter:1996cp}:
\begin{equation}
\partial_{t}\Gamma_{k}[h;\bar{g}]=\frac{1}{2}\textrm{Tr}\left(\Gamma_{k}^{(2;0)}[h;\bar{g}]+R_{k}[\bar{g}]\right)^{-1}\partial_{t}R_{k}[\bar{g}]+\textrm{ghost}\,,\label{RG_1}
\end{equation}
where the scale dependent, or running, effective action $\Gamma_{k}[h;\bar{g}]$
depends on the fluctuation and background metric. The ghost terms will
be discussed in a moment. Here $R_{k}[\bar{g}]$ is the cutoff kernel,
responsible for the regulation and for the coarse--graining. The
flow equation (\ref{RG_1}) can be derived by the RG improvement of
the regularized one--loop background gauge effective action, and as shown
in  \cite{Reuter:1996cp}, this improvement leads to an exact equation. The gauge
invariant part of the effective action is%
\footnote{We use $\Gamma_{k}[g]$ in place of the standard notation $\bar{\Gamma}_{k}[g]$
to simplify the notation. %
} $\Gamma_{k}[\bar{g}]\equiv\Gamma_{k}[0;\bar{g}]$ and satisfies a
flow equation given by setting $h_{\mu\nu}=0$ in (\ref{RG_1}). In
general the effective action can be written as:
\begin{equation}
\Gamma_{k}[h;\bar{g}]=\Gamma_{k}[\bar{g}+h]+S_{gf}[h;\bar{g}]\,,\label{RG_2}
\end{equation}
where we introduced the gauge--fixing action. The Hessian in (\ref{RG_1})
is then:
\begin{equation}
\Gamma_{k}^{(2;0)}[h;\bar{g}]=\Gamma_{k}^{(2)}[\bar{g}+h]+S_{gf}^{(2;0)}[h;\bar{g}]\label{RG_2.1}
\end{equation}
In this way, the flow equation for the gauge invariant part of the
effective action becomes:
\begin{eqnarray}
\partial_{t}\Gamma_{k}[\bar{g}] & = & \frac{1}{2}\textrm{Tr}\left(\Gamma_{k}^{(2)}[\bar{g}]+S_{gf}^{(2;0)}[0;\bar{g}]+R_{k}[\bar{g}]\right)^{-1}\partial_{t}R_{k}[\bar{g}]+\textrm{ghost}\,,\label{RG_3}
\end{eqnarray}
and will be used in the next sections to compute the beta functions
of $\Lambda_{k}$ and $G_{k}$. More specifically, to extract the
beta functions of a set of couplings $\lambda^{i}_k$, we expand both sides
of equation (\ref{RG_3}) on the relative operator basis $I_{i}[\bar{g}]$,
\begin{eqnarray}
\partial_{t}\Gamma_{k}[\bar{g}] & = & \sum_{i}\partial_{t}\lambda^{i}_k \, I_{i}[\bar{g}]\nonumber \\
\frac{1}{2}\textrm{Tr}\left(\Gamma_{k}^{(2)}[\bar{g}]+S_{gf}^{(2;0)}[0;\bar{g}]+R_{k}[\bar{g}]\right)^{-1}\partial_{t}R_{k}[\bar{g}] & = & \sum_{i}\beta^{i}I_{i}[\bar{g}]\,,\label{RG_3.1}
\end{eqnarray}
and by comparison we obtain the equations $\partial_{t}\lambda^{i}_k=\beta^{i}$,
i.e. the beta functions are the coefficients of the expansion of the
functional trace on the rhs of the flow equation. In the context of
quantum gravity, the expansion of the functional trace is performed
with the fundamental aid of the heat kernel expansion, in both its
local and non--local realizations \cite{Codello_Zanusso_2013}. These techniques
allow us to work covariantly at any step of the computations.

The ghost contribution in (\ref{RG_1}) or (\ref{RG_3}) depends on
the gauge; in CG we have:
\begin{equation}
\textrm{ghost}=-\frac{1}{2}\textrm{Tr}'\frac{\partial_{t}R_{k}(\Delta_{1})}{\Delta_{1}+R_{k}(\Delta_{1})}\,,\label{RG_4}
\end{equation}
where we used the fact that $L^{\dagger}L=\Delta_{1}$ and the excluded
zero modes are those of $L^{\dagger}$. In FG one instead finds:
\begin{equation}
\textrm{ghost}=-\textrm{Tr}\frac{\partial_{t}R_{k}(\Delta_{1})}{\Delta_{1}+R_{k}(\Delta_{1})}\,,\label{RG_5}
\end{equation}
still involving the spin one Laplacian but counted twice and with
no zero modes excluded. One of the virtues of equation (\ref{RG_3})
is that it holds in any dimension and allows, via the expansion (\ref{RG_3.1}),
the computation of the beta functions of any set of couplings. We
are going to exploit these properties in the next two sections, 4.3
and 4.4, to covariantly derive the RG flow of quantum gravity in,
respectively, $d=2$ and $d=2+\epsilon$.

\subsection{Quantum gravity in $d=2$}

In this Section we discuss the renormalization group flow in strictly
two dimensions. Since the invariant $\int\!\sqrt{g}R$
is topological it cannot be driving the RG flow, as instead does in
$d\geq2$, and another invariant must take its place in order to have
non--trivial beta functions. From the discussion of sections 2 and
3 we know that the natural candidate is the Polyakov action. Generally
we are then led to consider the following ansatz for the gauge invariant
part of the running effective action:
\begin{eqnarray}
\Gamma_{k}[g] & = & \int d^{2}x\,\sqrt{g}\left\{ \Lambda_{k}-\frac{1}{4\pi G_{k}}R-\frac{c_{k}}{96\pi}R\frac{1}{\Delta}R\right\} \nonumber \\
 & = & \Lambda_{k}I_{0}[g]-\frac{1}{4\pi G_{k}}I_{1}[g]+c_{k}S_{P}[g]\,.\label{CG_4}
\end{eqnarray}
Here $\Lambda_{k}$, $G_{k}$ and $c_{k}$ are running couplings,
the scale dependence of which contains the information about the RG
flow. We will soon see that the conformal anomaly does not renormalize,
at least within the set of operators that we are considering in (\ref{CG_4})
(see \cite{Codello_D'Odorico_Pagani_2014} for a deeper analysis of this
point) and we can thus set $c_{k}=c_{g}$ without any loss of generality.
We will also drop the bar over the background metric when
its presence is understood.

\subsubsection{Beta functions}

We will extract the beta functions for the couplings in (\ref{CG_4})
from the flow equation (\ref{RG_3}). The first thing we need to do
is to compute the Hessian of the action (\ref{CG_4}). To obtain the
quadratic action from which we can extract the Hessian, we need the
second variations of the operators $I_{0}[g]$, $I_{1}[g]$ and $S_{P}[g]$.
The details of these computations are given in the Appendix. We will
also employ the traceless--trace decomposition,
\begin{equation}
h_{\mu\nu}=\hat{h}_{\mu\nu}+\frac{1}{d}\bar{g}_{\mu\nu}h\qquad\qquad\bar{g}^{\mu\nu}\hat{h}_{\mu\nu}=0\,,\label{CG_4.01}
\end{equation}
in order both to simplify the second variations and to separate the
gauge part (traceless) from the physical part (trace). For $I_{0}[g]$
we have from (\ref{X_6}):
\begin{equation}
\delta^{2}I_{0}[g]=\int d^{2}x\sqrt{g}\left\{ \frac{\xi}{4}h^{2}+\frac{\xi-1}{2}\hat{h}^{\alpha\beta}\hat{h}_{\alpha\beta}\right\} \,,\label{CG_4.1}
\end{equation}
showing the in the standard parametrization we have only a traceless
contribution, while in the exponential parametrization we have only
a trace contribution. The second variation of $I_{1}[g]$ (\ref{X_10})
can be written as:
\begin{equation}
\delta^{2}I_{1}[g]=-\int d^{2}x\sqrt{g}\left\{ \frac{1}{2}\hat{h}^{\mu\nu}\left(\Delta+R\right)\hat{h}_{\mu\nu}+\hat{h}^{\mu\nu}\nabla_{\nu}\nabla_{\alpha}\hat{h}_{\mu}^{\alpha}\right\} \,,\label{CG_4.2}
\end{equation}
showing no dependence on $\xi$. We also note that this variation has
only a traceless part and will thus vanish when we employ the strict CG forcing
$\hat{h}_{\mu\nu}=0$, as expected from the
topological nature of the invariant. In a general CG with $\alpha\neq0$
we will see that the traceless contributions are clearly pure gauge.
In the Appendix we also report the details that lead to the second variation
of the Polyakov action (\ref{A_15}):
\begin{equation}
\delta^{2}S_{P}[g]=-\frac{1}{96\pi}\frac{1}{2}\int d^{2}x\,\sqrt{g}\left\{ h\left[\Delta-(\xi-1)R\right]h+2hA^{\mu\nu}\hat{h}_{\mu\nu}+\hat{h}_{\mu\nu}B_{\alpha\beta}^{\mu\nu}\hat{h}^{\alpha\beta}\right\} \,,\label{CG_5}
\end{equation}
where both $A^{\mu\nu}$ and $B_{\alpha\beta}^{\mu\nu}$ are known
tensors of which we don't need the explicit expression. Since they
are part of the traceless and traceless--trace sectors, in strict
CG they vanish, while, as before, in general CG they will be pure
gauge.

Before explicitly computing the Hessian we make an important point:
we recover the Liouville action (\ref{L_3.1}) only if we use the
exponential parametrization $\xi=1$. Using the first (\ref{A_11})
and second (\ref{CG_5}) variations of the Polyakov action we find
the following relation,
\begin{equation}
\delta S_{P}[g]+\frac{1}{2}\delta^{2}S_{P}[g]=-\frac{1}{24\pi}\int d^{2}x\sqrt{g}\left\{ \frac{h}{4}R+\frac{h}{4}\Delta\frac{h}{4}\right\} =-\frac{1}{24\pi}S_{L}\left[\frac{h}{4};g\right]\,,\label{CG_5.1}
\end{equation}
which we expected to hold given (\ref{CG_3}). We will argue that
this fact strongly supports the use of the exponential parametrization.

The gauge--fixing action (\ref{PI_1.3}) is purely traceless%
:
\[
S_{gf}[h;g]=\frac{1}{2\alpha}\int d^{2}x\sqrt{g}\,\hat{h}_{\mu\nu}\hat{h}^{\mu\nu}\,,
\]
and when added to the other variations (\ref{CG_4.1}), (\ref{CG_4.2})
and (\ref{CG_5}) gives the quadratic part of the action (\ref{CG_4}):
\[
\frac{1}{2}\Lambda_{k}\delta^{2}I_{0}[g]-\frac{1}{8\pi G_{k}}\delta^{2}I_{1}[g]+\frac{1}{2}\delta^{2}S_{P}[g]+S_{gf}[h;g]\qquad\qquad
\]
\begin{equation}
\qquad\qquad=\frac{1}{2}\int d^{2}x\,\sqrt{g}\left\{ h\Gamma h+2h\Gamma^{\mu\nu}\hat{h}_{\mu\nu}+\hat{h}_{\mu\nu}\left[\frac{1}{\alpha}\hat{\delta}_{\alpha\beta}^{\mu\nu}+\Gamma_{\alpha\beta}^{\mu\nu}\right]\hat{h}^{\alpha\beta}\right\} \,,\label{CG_5.11}
\end{equation}
where we defined the following tensors:
\begin{eqnarray}
\Gamma & = & \frac{Q^{2}}{8}\left[\Delta+(\xi-1)R+\frac{2\Lambda_{k}\xi}{Q^{2}}\right]\nonumber \\
\Gamma^{\mu\nu} & = & A^{\mu\nu}\nonumber \\
\Gamma_{\alpha\beta}^{\mu\nu} & = & \Lambda_{k}\frac{\xi-1}{2}\delta_{\alpha\beta}^{\mu\nu}+\frac{1}{4\pi G_{k}}\left[\frac{1}{2}\delta_{\alpha\beta}^{\mu\nu}(\Delta+R)+\nabla^{\nu}\nabla_{\alpha}\delta_{\beta}^{\mu}\right]+\frac{Q^{2}}{8}B_{\alpha\beta}^{\mu\nu}\,,\label{CG_5.12}
\end{eqnarray}
and $\hat{\delta}_{\alpha\beta}^{\mu\nu}$ is the identity in the
space of symmetric rank two tensors. We also conventionally define
\begin{equation}
Q=\sqrt{\frac{-c_{g}}{24\pi}}\,,\label{CG_5.121}
\end{equation}
preparing for discussing the connection with the Liouville theory approach
to two dimensional quantum gravity that we will make in a later Section.
Using the notation $\bar{\Gamma}$ to represent $\Gamma^{\mu\nu}$
and $\mathbf{\Gamma}$ to represent $\Gamma_{\alpha\beta}^{\mu\nu}$,
we can write and perform the multiplet trace implicit in the flow
equation (\ref{CG_4}):
\[
\textrm{tr}\left[\left(\begin{array}{cc}
\frac{1}{\alpha}\mathbf{1}+\mathbf{\Gamma}+\mathbf{R}_{k} & \bar{\Gamma}\\
\bar{\Gamma}^{T} & \Gamma+R_{k}
\end{array}\right)^{-1}\left(\begin{array}{cc}
\partial_{t}\mathbf{R}_{k} & 0\\
0 & \partial_{t}R_{k}
\end{array}\right)\right]=\frac{\partial_{t}R_{k}}{\Gamma+R_{k}}\qquad\qquad\qquad
\]
\begin{equation}
+\,\bar{\Gamma}^{T}\frac{\alpha}{\mathbf{1}+\alpha\left(\mathbf{\Gamma}+\mathbf{R}_{k}+\bar{\Gamma}^{T}\frac{1}{\Gamma+R_{k}}\bar{\Gamma}\right)}\bar{\Gamma}\frac{\partial_{t}R_{k}}{\Gamma+R_{k}}+\frac{\alpha\,\partial_{t}\mathbf{R}_{k}}{\mathbf{1}+\alpha\left(\mathbf{\Gamma}+\mathbf{R}_{k}+\bar{\Gamma}^{T}\frac{1}{\Gamma+R_{k}}\bar{\Gamma}\right)}\,.\label{CG_5.13}
\end{equation}
We see from (\ref{CG_5.13}) that after the inversion we can safely go to the strict CG
$\alpha=0$. Note also that equation (\ref{CG_5.13}) implicitly defines the tensor structure of the cutoff kernel.
When we insert (\ref{CG_5.13}) in the flow equation (\ref{RG_3}) we obtain the following form:
\begin{equation}
\partial_{t}\Gamma_{k}[g]=\frac{1}{2}\textrm{Tr}\frac{\partial_{t}R_{k}(\Delta_{0})}{\Delta_{0}+R_{k}(\Delta_{0})+\frac{2\Lambda_{k}\xi}{Q^{2}}}-\frac{1}{2}\textrm{Tr}'\frac{\partial_{t}R_{k}(\Delta_{1})}{\Delta_{1}+R_{k}(\Delta_{1})}\,,\label{CG_5.7}
\end{equation}
where we defined $\Delta_{0}=\Delta+(\xi-1)R$ as the spin zero operator
and rescaled the cutoff $R_{k}\rightarrow\frac{Q^{2}}{8}R_{k}$. This
is the flow equation for two dimensional quantum gravity in CG from
which now we will extract the beta function of $\Lambda_{k}$ and
$G_{k}$.

The functional traces of functions of Laplacian operators of the general
form $\Delta=-\nabla^{2}\mathbf{1}+\mathbf{U}$, like those in (\ref{CG_5.7}),
can be computed with the standard local heat kernel expansion:
\begin{equation}
\textrm{Tr}\,\mathbf{M}\, f(\Delta)=\frac{1}{(4\pi)^{d/2}}\sum_{n=0}^{\infty}Q_{\frac{d}{2}-n}[f]\int d^{d}x\,\textrm{tr}\left[\mathbf{M}\,\mathbf{b}_{2n}(\Delta)\right]\,,\label{HK_1}
\end{equation}
where $\mathbf{M}$ is a possible matrix insertion. The first two heat kernel coefficients, the only we will use,
are:
\begin{equation}
\mathbf{b}_{0}(\Delta)=\mathbf{1}\qquad\qquad\mathbf{b}_{2}(\Delta)=\mathbf{1}\frac{R}{6}-\mathbf{U}\,.\label{HK_2}
\end{equation}
The $Q$--functionals appearing in (\ref{HK_1}) are defined as $Q_{n}[f]=\frac{1}{\Gamma(n)}\int dz\, z^{n-1}f(z)$
if $n>0$ and as $Q_{n}[f]=(-1)^{|n|}f^{(|n|)}(0)$ if $n\leq0$ (see
the Appendix in \cite{Codello:2008vh} for more details).

We can now prove the non--renormalization of the anomaly following
\cite{Codello_2010}: the heat kernel expansion does not contain the invariant
$\int\sqrt{g}R\frac{1}{\Delta}R$ and thus the beta function of the anomaly coefficient is zero. As we will see in Section 4.4.2, only in the
$k\rightarrow0$ limit this operator will be produced. This justifies
the substitution $c_{k}\rightarrow c_{g}$ we previously made.

Using (\ref{HK_1}) we can immediately evaluate the traces in (\ref{CG_5.7})
\begin{eqnarray}
\textrm{Tr}\frac{\partial_{t}R_{k}(\Delta_{0})}{\Delta_{0}+R_{k}(\Delta_{0})+\frac{2\Lambda_{k}\xi}{Q^{2}}} & = & \frac{1}{4\pi}\int d^{2}x\,\sqrt{g}\left\{ Q_{1}\!\left[h_{k}\left(\frac{2\Lambda_{k}\xi}{Q^{2}}\right)\right]\, b_{0}(\Delta_{0})\right.\qquad\qquad\nonumber \\
 &  & \qquad\qquad\left.+Q_{0}\!\left[h_{k}\left(\frac{2\Lambda_{k}\xi}{Q^{2}}\right)\right]\, b_{2}(\Delta_{0})+O(R^{2})\right\} \,,\label{CG_5.8}
\end{eqnarray}
where the explicit values for the heat kernel coefficients are the
following:
\begin{equation}
b_{0}(\Delta_{0})=1\qquad\qquad b_{2}(\Delta_{0})=\frac{R}{6}+(1-\xi)R=\frac{7-6\xi}{6}R\,.\label{CG_6}
\end{equation}
We also introduced the notation $h_{k}(\omega)=\frac{\partial_{t}R_{k}}{z+R_{k}+\omega}$.
The ghost trace in (\ref{CG_5.7}) is:
\begin{equation}
\textrm{Tr}'\,\frac{\partial_{t}R_{k}(\Delta_{1})}{\Delta_{1}+R_{k}(\Delta_{1})}=\frac{1}{4\pi}\int d^{2}x\,\sqrt{g}\Big\{ Q_{1}[h_{k}(0)]\, b_{0}'(\Delta_{1})_{\mu}^{\mu}+Q_{0}[h_{k}(0)]\, b_{2}'(\Delta_{1})_{\mu}^{\mu}+O(R^{2})\Big\} \,,\label{CG_6.01}
\end{equation}
where the heat kernel coefficients with the zero mode extracted are%
\footnote{The zero modes of $\Delta_{1}\equiv L^{\dagger}L$ are those of $L^{\dagger}$;
one finds $N_{0}(L^{\dagger})=-3\chi$ for the the number of zero modes and thus the heat kernel coefficients
satisfy $b_{2}(\Delta_{1})=b_{2}'(\Delta_{1})+N_{0}(L^{\dagger})$.%
}:
\[
b_{0}'(\Delta_{1})_{\mu}^{\mu}=b_{0}(\Delta_{1})_{\mu}^{\mu}=\delta_{\mu}^{\mu}=2
\]
\begin{equation}
b_{2}'(\Delta_{1})_{\mu}^{\mu}=b_{2}(\Delta_{1})_{\mu}^{\mu}+3R=2\frac{R}{6}+R+3R=\frac{13}{3}R\,.\label{CG_6.1}
\end{equation}
The flow equation (\ref{CG_5.7}) then becomes:
\begin{eqnarray*}
\partial_{t}\Gamma_{k}[g] & = & \frac{1}{8\pi}\left\{ -Q_{1}\!\left[h_{k}\left(\frac{2\Lambda_{k}\xi}{Q^{2}}\right)\right]+2Q_{1}[h_{k}(0)]\right\} \int d^{2}x\sqrt{g}\\
 &  & \qquad\qquad+\frac{1}{8\pi}\left\{ \frac{7-6\xi}{6}Q_{0}\!\left[h_{k}\left(\frac{2\Lambda_{k}\xi}{Q^{2}}\right)\right]-\frac{26}{3}\right\} \int d^{2}x\sqrt{g}R+O(R^{2})\,,
\end{eqnarray*}
where we used the cutoff independent fact that $Q_{0}[h_{k}(0)]=2$. A comparison
with (\ref{CG_4}) gives the beta functions:
\begin{eqnarray}
\partial_{t}\Lambda_{k} & = & \frac{1}{8\pi}\left\{ -Q_{1}\!\left[h_{k}\left(\frac{2\Lambda_{k}\xi}{Q^{2}}\right)\right]+2Q_{1}[h_{k}(0)]\right\} \nonumber \\
\partial_{t}\left(-\frac{1}{G_{k}}\right) & = & \frac{1}{2}\left\{ \frac{7-6\xi}{6}Q_{0}\!\left[h_{k}\left(\frac{2\Lambda_{k}\xi}{Q^{2}}\right)\right]-\frac{26}{3}\right\} \,.\label{CG_6.2}
\end{eqnarray}
In the case $\Lambda_{k}=0$ this last relation becomes:
\begin{equation}
\partial_{t}\left(-\frac{1}{G_{k}}\right)=\underbrace{\frac{7-6\xi}{6}}_{\textrm{spin 0}}\underbrace{-\frac{26}{6}}_{\textrm{ghost}}=-\frac{19+6\xi}{6}\,,\label{CG_6.3}
\end{equation}
or more explicitly:
\begin{equation}
\partial_{t}G_{k}=\left\{ \begin{array}{c}
-\frac{1}{6}19G_{k}^{2}\qquad\xi=0\\
-\frac{1}{6}25G_{k}^{2}\qquad\xi=1
\end{array}\right.\,,\label{CG_7}
\end{equation}
which shows that the gravitational contribution to the beta function of
Newton's constant is $-19$ in the standard parametrization, while
it is $-25$ in the exponential parametrization. This result is new
and shows that there is a dependence on the field parametrization,
at least in the computation we have done, also in strictly two dimensional quantum
gravity. Later we will compare this with the relative computation in $d=2+\epsilon$. 
Using now the relation (\ref{M_11}) between $c_{g}$ and the
beta function of Newton's constant leads to the following
value for the total gravitational anomaly coefficient:
\begin{equation}
c_{g}=\underbrace{c_{\Phi}}_{\textrm{matter}}\:\underbrace{+7-6\xi}_{\textrm{spin 0}}\:\underbrace{-26}_{\textrm{ghost}}=\left\{ \begin{array}{c}
c_{\Phi}-19\qquad\xi=0\\
c_{\Phi}-25\qquad\xi=1
\end{array}\right.\,,\label{CG_8.1}
\end{equation}
where we made the break--down of the various contributions and added
the matter contribution. This is, in the $\xi=1$ case, the result
that we pre--announced in Section 3. As we see, the ghost
contribute the famous $-26$, while the trace spin zero part of the
metric contributes like a standard scalar in the exponential parametrization
and like seven scalars in the standard parametrization. From many other
computations and constructions (see \cite{Ambjorn:1997di} for a comprehensive discussion of the literature) we
know that the correct value is the one found in the exponential parametrization.
Why the standard parametrization fails is to be understood. We note
finally that this values are scheme independent since their derivation relied
only the fact that $Q_{0}[h_{k}(0)]=2$, which is true for any admissible cutoff shape.

In the general case $\Lambda_{k}\neq0$ we find, employing Litim's
cutoff, i.e. $R_k(z)=(k^2-z)\theta(k^2-z)$, the following beta functions:
\begin{eqnarray}
\partial_{t}\tilde{\Lambda}_{k} & = & -2\tilde{\Lambda}_{k}+\frac{1}{4\pi}\left\{ \frac{1}{1+\frac{2\tilde{\Lambda}_{k}\xi}{Q^{2}}}-2\right\} \nonumber \\
\partial_{t}G_{k} & = & \left\{ \frac{7-6\xi}{6}\frac{1}{1+\frac{2\tilde{\Lambda}_{k}\xi}{Q^{2}}}-\frac{26}{6}\right\} G_{k}^{2}\,.\label{CG_9}
\end{eqnarray}
These beta functions show that two dimensional quantum gravity is
asymptotically free. We refer to the literature for more on this point \cite{Codello:2008vh}.

\subsubsection{Scaling exponents}

Having computed $c_{g}$ we now turn to the computation of the scaling
exponents $\alpha,\beta,...$ in this way completing the determination
of the scaling laws of Section 3. 

The area operator scales classically as $I_{0}[\lambda g]=\int\sqrt{\textrm{det}\left[\lambda g\right]}=\lambda I_{0}[g]$.
Fluctuations will generically change this by adding a nontrivial anomalous dimension
$I_{0}[\lambda g]=\lambda^{\alpha}I_{0}[g]$. To account for this we will consider
the generalized composite operator:
\begin{equation}
I_{0}[g]=\int d^{2}x\sqrt{g}^{\alpha}\,,\label{SE_1}
\end{equation}
where $\alpha$ is the scaling exponent we want to determine. 
One can attribute a dimensionality either to the coordinates or to the metric.
Let's assume the last case, so $[g_{\mu\nu}]=k^{-2}$. The exponent
$\alpha$ is determined self--consistently by requiring that the operator
dimension of the area operator $I_{0}[g]$ is two. In this section
we will work strictly in CG and we will employ the exponential parametrization
$\xi=1$ only, since this is the one that leads to the correct scaling
exponents.

To determine the anomalous scaling dimension of $I_{0}[g]$ we add
to the effective action (\ref{CG_4}) the term%
\footnote{Remember that composite operators are introduced as $\int\mathcal{D}\phi\, e^{-S[\phi]+\int J\mathcal{O}(\phi)}$.
In any case the the sign in front of $Z_{k}$ drops out from the final
formula for the scaling dimension.%
} $-Z_{k}\int\sqrt{g}^{\alpha}$, where $Z_{k}$ is the wave--function
renormalization constant of $I_{0}[g]$. The Hessian to insert in
the flow equation (\ref{RG_1}) (note that we are considering the
full bi--metric action $\Gamma_{k}[h;g]$) is of the general form
(\ref{CG_5.12}). Since we work in CG we need only
the trace part of the Hessian, which now reads:
\begin{equation}
\Gamma=\frac{Q^{2}}{8}\Delta-\frac{\alpha^{2}}{4}Z_{k}e^{\frac{\alpha}{2}h}\,,\label{SE_2}
\end{equation}
where we remember $Q=\frac{-c_{g}}{24\pi}=\frac{25-c_{\Phi}}{24\pi}$.
Using (\ref{SE_2}) in the flow equation (\ref{RG_1}) after performing a rescaling
the cutoff $R_{k}\rightarrow\frac{Q^{2}}{8}R_{k}$ as before, we find
the following expression:
\begin{eqnarray}
\partial_{t}\Gamma_{k}[h;g] & = & \frac{1}{2}\textrm{Tr}\frac{\frac{Q^{2}}{8}\partial_{t}R_{k}(\Delta)}{\frac{Q^{2}}{8}\Delta-\frac{\alpha^{2}}{4}Z_{k}e^{\frac{\alpha}{2}h}+\frac{Q^{2}}{8}R_{k}(\Delta)}\nonumber \\
 & = & \frac{1}{2}\textrm{Tr}\frac{\partial_{t}R_{k}(\Delta)}{\Delta+R_{k}(\Delta)}+\frac{\alpha^{2}}{Q^{2}}Z_{k}\textrm{Tr}\left\{ \frac{\partial_{t}R_{k}(\Delta)}{[\Delta+R_{k}(\Delta)]^{2}}e^{\frac{\alpha}{2}h}\right\} +...\label{SE_3}
\end{eqnarray}
Without loss of generality we can set the background metric to be
the flat metric in order to simplify the computations. The lhs of
(\ref{SE_3}) is then $\partial_{t}\Gamma_{k}[h;\delta]=-\left(\int d^{2}x\right)\,\partial_{t}Z_{k}e^{\frac{\alpha}{2}h}$.
Using the heat kernel expansion (\ref{HK_1}) to its lowest order
to evaluate the functional trace and comparing both sides leads to
the following equation for the wave--function renormalization:
\begin{equation}
\partial_{t}Z_{k}=-\frac{\alpha^{2}}{Q^{2}}Z_{k}\frac{1}{4\pi}Q_{1}[g_{k}]\,.\label{SE_3.1}
\end{equation}
where $g_{k}(z)=\frac{\partial_{t}R_{k}(z)}{[z+R_{k}(z)]^{2}}$. The
$Q$--functional in (\ref{SE_3.1}) is scheme independent $Q_{1}[g_{k}]=2$
and thus we find:
\begin{equation}
\partial_{t}Z_{k}=-\frac{\alpha^{2}}{2\pi Q^{2}}Z_{k}\,.\label{SE_4}
\end{equation}
In terms of dimensionless variables, $\left[\sqrt{g}^{\alpha}\right]=k^{-2\alpha}$
and $[Z_{k}]=k^{2\alpha-\eta}$, where
\begin{equation}
\eta=-\partial_{t}\log Z_{k}=\frac{\alpha^{2}}{2\pi Q^{2}}\,,\label{SE_4.1}
\end{equation}
is the anomalous dimension of the operator $I_{0}[g]$. The request
that this operator scales like an area leads to $2\alpha-\eta=2$,
or more explicitly using (\ref{SE_4.1}):
\begin{equation}
2\alpha-\frac{\alpha^{2}}{2\pi Q^{2}}=2\,,\label{SE_5}
\end{equation}
which, knowing that $Q=\frac{25-c_{\Phi}}{24\pi}$, is equivalent
to:
\begin{equation}
2-2\alpha+2\frac{6}{25-c_{\Phi}}\alpha^{2}=0\,.\label{SE_7}
\end{equation}
The solution to (\ref{SE_7}) is:
\begin{equation}
\alpha=\frac{25-c_{\Phi}-\sqrt{(1-c_{\Phi})(25-c_{\Phi})}}{12}\,,\label{SE_8}
\end{equation}
where we picked the branch that leads to $\alpha\rightarrow1$ in
the classical limit $c_{\Phi}\rightarrow-\infty$.

This can now be applied to a general composite operator $I_{\mathcal{O}}[\Phi;g]=\int\!\sqrt{g}\,\mathcal{O}[\Phi;g]$,
the only difference will be the classical scaling dimension, that
we generally call $\Delta_{0}$ (we had $\Delta_{0}=0$ in the case
of $I_{0}[g]$), and equation (\ref{SE_7}) generalizes to:
\begin{equation}
1-\Delta_{0}-\beta+\frac{6}{25-c_{\Phi}}\beta^{2}=0\,.\label{SE_9}
\end{equation}
This immediately gives the scaling exponent $\beta$ as
\begin{equation}
\beta=\frac{25-c_{\Phi}-\sqrt{(1+24\Delta_{0}-c_{\Phi})(25-c_{\Phi})}}{12}\,.\label{SE_10}
\end{equation}
This is the scaling that enters in the KPZ relation. To find the alternative
form of this scaling, simply use $\beta=\alpha(1-\Delta)$ in the
equation that defines $\beta$, to find
\begin{eqnarray}
\Delta-\Delta_{0} & = & 1-\Delta_{0}-(1-\Delta)\nonumber \\
 & = & (1-\Delta)\left[-1+\alpha-\frac{6\alpha^{2}}{25-c_{\Phi}}(1-\Delta)\right]\nonumber \\
 & = & \frac{6\alpha^{2}}{25-c_{\Phi}}\Delta(1-\Delta)\,.\label{SE_11}
\end{eqnarray}
Likewise, the exponent $\tilde{\alpha}$ needed in the scaling of
the geodesic distance is determined as the scaling of the Laplacian.
Since in $d=2$ the operator $\sqrt{g}\Delta$ is scale--invariant,
the classical scaling of the Laplacian is fixed to be that of an inverse
area, and is thus found by solving
\begin{equation}
-1-\tilde{\alpha}+\frac{6}{25-c_{\Phi}}\tilde{\alpha}^{2}=0\,,\label{SE_12}
\end{equation}
which gives
\begin{equation}
\tilde{\alpha}=\frac{25-c_{\Phi}-\sqrt{(49-c_{\Phi})(25-c_{\Phi})}}{12}\,,\label{SE_13}
\end{equation}
another relation that we used before.

\subsubsection{Connection with Liouville theory}

In this Section we want to make contact with the standard way to determine
$c_{g}$ which is via Liouville theory \cite{Distler_Kawai_1988,David_1988,Kawai_Ninomiya_1990}. In CG the partition
function, in terms of the standard non--Weyl invariant measure, takes the form:
\begin{equation}
Z=\int\mathcal{D}_{g}\hat{h}_{\mu\nu}\mathcal{D}_{g}h\,\delta[\hat{h}_{\mu\nu}]Z_{\Phi}[g]Z_{gh}[g]\,,\label{LT_1}
\end{equation}
where $g_{\mu\nu}=\bar{g}_{\mu\lambda}e^{h_{\;\nu}^{\lambda}}$. The
integral over traceless metric fluctuations $\hat{h}_{\mu\nu}=h_{\mu\nu}-\frac{1}{2}g_{\mu\nu}h$
can be performed directly, imposing in this way the CG gauge--fixing
condition strictly. The partition function then becomes Gaussian:
\begin{equation}
Z=\int\mathcal{D}_{\bar{g}e^{h/2}}h\, Z_{\Phi}[\bar{g}e^{h/2}]Z_{gh}[\bar{g}e^{h/2}]\,,\label{LT_2}
\end{equation}
where we already know the matter and ghost partition functions $Z_{\Phi}$
and $Z_{gh}$ as a function of $g$. Under a Weyl rescaling they transform
as follows:
\begin{align}
Z_{\Phi}[\bar{g}e^{h/2}] & =Z_{\Phi}[\bar{g}]e^{\frac{c_{\Phi}}{24\pi}S_{L}[h/4;\bar{g}]}\nonumber \\
Z_{gh}[\bar{g}e^{h/2}] & =Z_{gh}[\bar{g}]e^{-\frac{26}{24\pi}S_{L}[h/4;\bar{g}]}\,.\label{LT_3}
\end{align}
Assuming also that $h$ behaves as a standard scalar of weight zero,
so that its measure is subject to (\ref{M_1}), leads to the following
form for the partition function:
\begin{equation}
Z=Z_{\Phi}[\bar{g}]Z_{gh}[\bar{g}]\int\mathcal{D}_{\bar{g}}h\, e^{\frac{c_{\Phi}+c_{h}-26}{24\pi}S_{L}[h/4;\bar{g}]}\,,\label{LT_4}
\end{equation}
where $c_{h}$ is in principle unknown and will be fixed in a moment
(even if we expect it to be one). We can now work with the more standard
Liouville variable $\sigma=\frac{h}{4}$. Since $Z$ is not affected
by a Weyl rescaling we must check that the same is true for the rhs
of (\ref{LT_4}), thus it must be independent of the shift $\bar{g}\rightarrow\bar{g}e^{2\chi}$
and $\sigma\rightarrow\sigma-\chi$ if the conformal factor measure
is translation invariant $\mathcal{D}_{\bar{g}}(\sigma-\chi)=\mathcal{D}_{\bar{g}}\sigma$.
We have:
\begin{align}
S_{L}[\sigma-\chi;\bar{g}e^{2\chi}] & =\int d^{2}x\sqrt{\bar{g}}e^{2\chi}\left[2(\sigma-\chi)e^{-2\chi}\bar{\Delta}(\sigma-\chi)+2(\sigma-\chi)e^{-2\chi}(\bar{R}+2\bar{\Delta}\chi)\right]\nonumber \\
 & =S_{L}[\sigma;\bar{g}]-S_{L}[\chi;\bar{g}]\,,\label{L_4.1}
\end{align}
so we find that indeed:
\[
\mathcal{D}_{\bar{g}e^{2\chi}}\sigma\, Z_{\Phi}[\bar{g}e^{2\chi}]Z_{gh}[\bar{g}e^{2\chi}]e^{\frac{c_{\Phi}+c_{h}-26}{24\pi}S_{L}[\sigma-\chi;\bar{g}e^{2\chi}]}=\mathcal{D}_{\bar{g}}\sigma\, Z_{\Phi}[\bar{g}]Z_{gh}[\bar{g}]e^{\frac{c_{\Phi}+c_{h}-26}{24\pi}S_{L}[\sigma;\bar{g}]}\,.
\]
Thus (\ref{LT_4}) is well defined. We define conventionally $Q=\sqrt{\frac{26-c_{\Phi}-c_{h}}{24\pi}}$,
implicitly assuming that $c_{\Phi}+c_{h}<26$ in order to a have a
positive definite action. We need to evaluate the following Gaussian integral:
\begin{equation}
I=\int\mathcal{D}_{g}\sigma\, e^{-Q^{2}\int\sqrt{g}\left(\sigma\Delta\sigma+\sigma R\right)}\,.\label{LT_5}
\end{equation}
This is easily performed by just completing the square:
\begin{equation}
\frac{1}{2}\int d^{2}x\sqrt{g}\left(\sigma+\frac{1}{\Delta}R\right)\Delta\left(\sigma+\frac{1}{\Delta}R\right)=\int d^{2}x\sqrt{g}\left[\frac{1}{2}\sigma\Delta\sigma+\sigma R+\frac{1}{2}R\frac{1}{\Delta}R\right]\,,\label{LT_6}
\end{equation}
where we integrated by parts. Inserting (\ref{LT_6}) into (\ref{LT_5})
and shifting the integration variable as $\sigma\rightarrow Q\left(\sigma+\frac{1}{\Delta}R\right)$
gives:
\begin{align}
I & =e^{\frac{Q^{2}}{2}\int\sqrt{g}R\frac{1}{\Delta}R}\int\mathcal{D}_{g}\sigma\, e^{-\frac{1}{2}\int\sqrt{g}\sigma\Delta\sigma}\nonumber \\
 & =e^{\frac{Q^{2}}{2}\int\sqrt{g}R\frac{1}{\Delta}R}e^{-\frac{1}{2}\textrm{Tr}\log\Delta}\nonumber \\
 & =e^{\frac{1+48\pi Q^{2}}{96\pi}\int\sqrt{g}R\frac{1}{\Delta}R}\,,\label{LT_7}
\end{align}
where we evaluated the Gaussian integral and we collected all terms.
Incidentally this shows the Liouville action has central charge $c_{L}=1+48\pi Q^{2}$.


Using (\ref{LT_7}) in (\ref{LT_4}) and remembering the form of $Z_{m}[\bar{g}]$
and of $Z_{gh}[\bar{g}]$ gives:
\begin{equation}
Z=e^{\frac{c_{\Phi}-26}{96\pi}\int\sqrt{g}R\frac{1}{\Delta}R}e^{\frac{1+26-c_{\Phi}-c_{h}}{96\pi}\int d^{2}x\sqrt{g}R\frac{1}{\Delta}R}=e^{\frac{1-c_{h}}{96\pi}\int d^{2}x\sqrt{g}R\frac{1}{\Delta}R}\,.\label{LT_8}
\end{equation}
Demanding that $Z$ is Weyl invariant leads to $c_{h}=1$ since the
only (on--shell) effective action which is Weyl invariant in $d=2$
is $\Gamma[g]=0$, which in turn implies $Z=1$. Said in an equivalent
way, to ask for $c_{h}=1$ is equivalent to ask that conformal invariance
is restored. This was basically the original argument of \cite{Distler_Kawai_1988,David_1988};
see also \cite{Mottola:1995sj}. We then have:
\begin{equation}
Q=\sqrt{\frac{25-c_{\Phi}}{24\pi}}\,,\label{LT_10}
\end{equation}
which establishes again the fundamental result $c_{g}=c_{\Phi}-25$
and agrees with our previous definitions. In the context of Liouville
theory the exponents $\alpha,\beta,...$ are now related to the scaling
dimensions of the so--called vertex operators, $V_{\alpha}=e^{\alpha\sigma}$,
and can be computed by standard CFT methods (we refer to \cite{Mottola:1995sj}
for more details). Needless to say, the results are the
same as those we gave in the previous sections. With the knowledge
of $c_{g}$ and the scaling exponents one then derives the scaling
relations for various observables within Liouville theory and recovers
the results we presented in Section 3. We want to remark that our
derivation shows that is possible to respect covariance and that Liouville
theory is not the only way, or the fundamental way, to establish scaling
relations in the continuum. It is just one way to perform the analysis;
more precisely, is the way to exploit the fact that in CG quantum
gravity in $d=2$ is a Gaussian theory; in fact all results of Liouville
theory derive from the use of the Gaussian integral (\ref{LT_7})
or generalizations to include vertex operators, i.e. the integral
(\ref{LT_7}) in presence of external currents.

\subsection{Quantum gravity in $d=2+\epsilon$ }

In this section we consider quantum gravity in dimension greater than
two. This is the case considered in almost all studies of renormalization in the context of quantum
gravity, starting from \cite{key-7} to the works of \cite{Kawai_Ninomiya_1990,Kawai_Kitazawa_Ninomiya_1992}
and later to the studies of gravity in the context of asymptotic safety
\cite{Reuter:2011ah,Codello:2008vh}. A recent study of the dependence of the beta functions on
 field parametrizations in this last context as been presented recently \cite{Nink_2014};
 for an application in the context of unimodular quantum gravity see \cite{Eichhorn:2013xr}.

Two things happens in $d>2$: the Polyakov action is not anymore induced
by matter and ghost fluctuations and the operator $\int\!\sqrt{g}R$
ceases to be topological. Somehow the latter start playing the role
of the former, but we cannot in general expect the arguments and scaling
relations of Section 2 and 3 to still be valid, since they were genuine
to $d=2$. In any case, we may hope to obtain a continuous $\epsilon\rightarrow0$
limit by employing $\int\!\sqrt{g}R$ in place of $S_{P}[g]$. As we
will explain more precisely later in this Section, this requires a
careful choice of the regulator in order to suppress the pathological
limit the Hessan of $\int\!\sqrt{g}R$ has when $d\rightarrow2$. We
will also see that, when we consider the finite part of the effective
action, only within the exponential parametrization we will be able
to take the limit $\epsilon\rightarrow0$. We thus consider the following
ansatz for the gauge invariant part of the effective action:
%
\begin{eqnarray}
\Gamma_{k}[g] & = & \Lambda_{k}\int d^{d}x\sqrt{g}-\frac{1}{4\pi G_{k}}\int d^{d}x\sqrt{g}R\nonumber \\
 & = & \Lambda_{k}I_{0}[g]-\frac{1}{4\pi G_{k}}I_{1}[g]\,,\label{FG_1}
\end{eqnarray}
and compute the beta functions of the cosmological constant and of
Newton's constant from those terms proportional to the invariants
$I_{0}[g]$ and $I_{1}[g]$ stemming from the expansion of functional
traces on the rhs of the flow equation (\ref{RG_1}).

\subsubsection{Beta functions}

As in Section 4.3.1, to compute the Hessian needed in the RG flow equation
(\ref{RG_3}) we first derive the quadratic action. We will consider
only the gauge $\alpha=1$ since this allows us to employ heat kernel
methods to compute the functional traces. We also employ the traceless--trace
decomposition (\ref{CG_4.01}). The second variation of $I_{0}[g]$
in arbitrary dimension is given in equation (\ref{X_6}) of the Appendix and reads:
\begin{equation}
\frac{1}{2}\Lambda_{k}\delta^{2}I_{0}[g]=\frac{1}{2}\Lambda_{k}\int d^{d}x\sqrt{g}\left(\frac{d-2+2\xi}{4d}h^{2}+\frac{\xi-1}{2}\hat{h}^{\alpha\beta}\hat{h}_{\alpha\beta}\right)\,;\label{FG_2}
\end{equation}
while the second variation of $I_{1}[g]$, when summed to the FG
gauge--fixing action (\ref{PI_1.4}) and evaluated on a spherical
background, is given in equation (\ref{X_13}) of the Appendix, or:
\[
-\frac{1}{2}\frac{1}{4\pi G_{k}}\delta^{2}I_{1}[g]+S_{gf}[h;g]=\frac{1}{2}\int d^{d}x\sqrt{g}\left\{ \frac{1}{2}\hat{h}^{\mu\nu}\left(\Delta+\frac{d^{2}-3d+4}{d(d-1)}R-\frac{d-2}{d}\xi R\right)\hat{h}_{\mu\nu}\right.
\]
\begin{equation}
\qquad\qquad\qquad\qquad\left.-\frac{d-2}{4d}h\left(\Delta+\frac{d-4+2\xi}{d}R\right)h\right\} \,.\label{FG_5}
\end{equation}
We will perform the replacement \cite{Codello:2013fpa}:
\begin{equation}
h_{\mu\nu}\rightarrow\sqrt{8\pi G_{k}}h_{\mu\nu}\,,\label{FG_3}
\end{equation}
in the expansions (\ref{FG_2}--\ref{FG_5}) and rescale $\Lambda_k$ appropriately. We can define the symmetric spin
two tensor identity $\delta_{\rho\sigma}^{\mu\nu}=\frac{1}{2}\left(\delta_{\rho}^{\mu}\delta_{\sigma}^{\nu}+\delta_{\sigma}^{\mu}\delta_{\rho}^{\nu}\right)$
and the trace projector $P_{\rho\sigma}^{\mu\nu}=\frac{1}{d}g^{\mu\nu}g_{\rho\sigma}$;
then we have:
\begin{eqnarray}
\hat{h}_{\alpha\beta} & = & h_{\alpha\beta}-\frac{1}{2}g_{\alpha\beta}h=\left(\delta_{\alpha\beta}^{\mu\nu}-\frac{1}{2}g_{\alpha\beta}g^{\mu\nu}\right)h_{\mu\nu}\equiv(\mathbf{1}-\mathbf{P})_{\alpha\beta}^{\mu\nu}h_{\mu\nu}\nonumber \\
\frac{1}{2}g_{\alpha\beta}h & = & \frac{1}{2}g_{\alpha\beta}g^{\mu\nu}h_{\mu\nu}\equiv\mathbf{P}_{\alpha\beta}^{\mu\nu}h_{\mu\nu}\,.\label{CG_5.3}
\end{eqnarray}
Note that $\mathbf{1}-\mathbf{P}$ and $\mathbf{P}$ are orthogonal
projectors into the trace and traceless subspaces in the space of
symmetric tensors. In terms of these projectors we can now write
the gravitational Hessian in the following way:
%
%
%
\begin{equation}
\mathbf{\Gamma}_{k}^{(2;0)}[0;g]=\frac{1}{2}(\mathbf{1-P})\left[\Delta_{2}+(\xi-1)\Lambda_{k}\right]-\frac{d-2}{4}\mathbf{P}\left[\Delta_{0}-\frac{d-2+2\xi}{d(d-2)}\Lambda_{k}\right]\,,\label{FG_6}
\end{equation}
where we defined the spin two and spin zero differential operators:
\begin{eqnarray}
\Delta_{2} & = & \Delta+\left(\frac{d^{2}-3d+4}{d(d-1)}-\frac{d-2}{d}\xi\right)R\nonumber \\
\Delta_{0} & = & \Delta+\frac{d-4+2\xi}{d}R\,,\label{FG_7}
\end{eqnarray}
while the ghost Hessian is the spin one differential operator $\Delta_1$ is given in (\ref{PI_8}) .
%
%
We need now to choose the cutoff kernel, the structure of the inverse
propagator (\ref{FG_6}) suggest the following:
\begin{equation}
\mathbf{R}_{k}=(\mathbf{1}-\mathbf{P})R_{k}(\mathbf{\Delta}_{2})-\frac{d-2}{2}\mathbf{P}R_{k}(\Delta_{0})\,.\label{FG_10}
\end{equation}
This natural choice is actually non--trivial and is ultimately responsible
for the continuity, that will discuss in a moment, of the $\epsilon\rightarrow0$
limit and to the taming of the ``wrong'' sign of the spin zero inverse
propagator. It was first introduced by \cite{Kawai_Ninomiya_1990,Kawai_Kitazawa_Ninomiya_1992} and later proposed in
the context we are considering by \cite{Reuter:1996cp}. It is easy now to write
down explicitly the full regularized graviton propagator:
\[
\left[(\mathbf{1-P})\Big(\mathbf{\Delta}_{2}+R_{k}(\mathbf{\Delta}_{2})+(\xi-1)\Lambda_{k}\Big)-\frac{d-2}{2}\mathbf{P}\left(\Delta_{0}+R_{k}(\Delta_{0})-\frac{d-2+2\xi}{d(d-2)}\Lambda_{k}\right)\right]^{-1}=\qquad\qquad
\]
\begin{equation}
\qquad\qquad=\;(\mathbf{1}-\mathbf{P})\frac{1}{\mathbf{\Delta}_{2}+R_{k}(\mathbf{\Delta}_{2})+(\xi-1)\Lambda_{k}}-\frac{2}{d-2}\mathbf{P}\frac{1}{\Delta_{0}+R_{k}(\Delta_{0})-\frac{d-2+2\xi}{d(d-2)}\Lambda_{k}}\,.\label{FG_11}
\end{equation}
%
%
Now when we multiply (\ref{FG_11}) with $\partial_{t}\mathbf{R}_{k}$
both the $d=2$ pole and the minus sign disappear and the flow equation,
and lately the beta functions, will not suffer of these problems in
the $\epsilon\rightarrow0$ limit, problems that are related to the
topological nature of the invariant $\int\!\sqrt{g}R$ when $d=2$.
As said, this choice of regulator is the one responsible for the good
behaviour of the $\epsilon\rightarrow0$ limit.

To proceed, we insert in the graviton part of the flow equation the
identity in the space of symmetric rank two tensor in the form $\mathbf{1}=(\mathbf{1}-\mathbf{P})+\mathbf{P}$.
After adding the ghost contribution (\ref{RG_5}) this gives the following flow equation:%
%
\begin{eqnarray}
\partial_{t}\Gamma_{k}[g] & = & \frac{1}{2}\textrm{Tr}\,(\mathbf{1}-\mathbf{P})\frac{\partial_{t}R_{k}(\Delta_{2})}{\mathbf{\Delta}_{2}+R_{k}(\Delta_{2})+(\xi-1)\Lambda_{k}}+\frac{1}{2}\textrm{Tr}\,\mathbf{P}\frac{\partial_{t}R_{k}(\Delta_{0})}{\Delta_{0}+R_{k}(\Delta_{0})-\frac{d-2+2\xi}{d(d-2)}}\nonumber \\
 &  & -\textrm{Tr}\,\delta_{\nu}^{\mu}\frac{\partial_{t}R_{k}(\Delta_{1})}{\Delta_{1}+R_{k}(\Delta_{1})}\,.\label{FG_14}
\end{eqnarray}
It is now easy to evaluate the traces using the local heat kernel
expansion (\ref{HK_1}). We find the following heat kernel coefficients:
\[
\textrm{tr}\left[(\mathbf{1}-\mathbf{P})\mathbf{b}_{2}(\mathbf{\Delta}_{2})\right]=\frac{d^{2}+d-2}{2}\left[\frac{R}{6}-\left(\frac{d^{2}-3d+4}{d(d-1)}-\frac{d-2}{2d}\xi\right)R\right]\underset{d=2}{=}-\frac{5}{3}R\,,
\]
where we used $\textrm{tr}\,(\mathbf{1}-\mathbf{P})=\frac{d^{2}+d-2}{2}$;
\[
\textrm{tr}\left[\mathbf{P}\mathbf{b}_{2}(\Delta_{0})\right]=\frac{R}{6}-\frac{d-4+2\xi}{d}R\underset{d=2}{=}\frac{7-6\xi}{6}R\,;
\]
showing that only the conformal mode is $\xi$--dependent;
\[
\textrm{tr}\, b_{2}(\Delta_{1})=\delta_{\mu}^{\mu}\frac{R}{6}+R_{\mu}^{\mu}=\frac{d+6}{6}R\underset{d=2}{=}\frac{4}{3}R\,;
\]
showing that there is nothing universal in the ghost trace.
Then, to linear order in the curvature and for $d=2$, the flow equation (\ref{FG_14}) is:
\begin{eqnarray}
\partial_{t}\Gamma_{k}[g]\!\! & = & \!\!\frac{1}{8\pi}\left\{ 2Q_{1}\!\left[h_{k}\Big((\xi-1)\Lambda_{k}\Big)\right]+Q_{1}\!\left[h_{k}\!\left(-\frac{\xi}{\epsilon}\Lambda_{k}\right)\right]-2Q_{1}\!\left[h_{k}\left(0\right)\right] \right\}\! \int d^{2}x\sqrt{g}\nonumber \\
 &  & \!\!+\frac{1}{8\pi}\left\{ -\frac{5}{3}Q_{0}\!\left[h_{k}\!\Big((\xi-1)\Lambda_{k}\Big)\right]+\frac{7-6\xi}{6}Q_{0}\!\left[h_{k}\!\left(-\frac{\xi}{\epsilon}\Lambda_{k}\right)\right]-\frac{8}{3}Q_{0}[h_{k}(0)]\right\} \!\int d^{2}x\sqrt{g}R\nonumber \\
 &  & +\,O(R^{2})\,.\label{FG_15}
\end{eqnarray}
%
%
This leads to the following beta functions:
\begin{eqnarray}
\partial_{t}\Lambda_{k} \!& = &\! \frac{1}{8\pi}\left\{ 2Q_{1}\!\left[h_{k}\Big((\xi-1)\Lambda_{k}\Big)\right]+Q_{1}\!\left[h_{k}\!\left(\!-\frac{\xi}{\epsilon}\Lambda_{k}\right)\right]-2Q_{1}\!\left[h_{k}\left(0\right)\right]  \right\} \nonumber \\
\partial_{t}\left(-\frac{1}{G_{k}}\right)\! & = &\! \frac{1}{2} \left\{ -\frac{5}{3}Q_{0}\!\left[h_{k}\!\Big((\xi-1)\Lambda_{k}\Big)\right]+\frac{7-6\xi}{6}Q_{0}\!\left[h_{k}\!\left(\!-\frac{\xi}{\epsilon}\Lambda_{k}\right)\right]-\frac{8}{3}Q_{0}[h_{k}(0)]\right\} \,.\qquad
\label{FG_16}
\end{eqnarray}
%
%
%
In the case $\Lambda_{k}=0$ we find, as in the previous section,
the universal beta function for Newton's constant (using the scheme independent value $Q_{0}[h_{k}(0)]=2$):
\begin{equation}
\partial_{t}\left(-\frac{1}{G_{k}}\right)=\underbrace{-\frac{5}{3}}_{\textrm{spin 2}}\underbrace{+\frac{7-6\xi}{6}}_{\textrm{spin 0}}\underbrace{-\frac{8}{3}}_{\textrm{ghost}}=-\frac{19+6\xi}{6}\,,\label{FG_17}
\end{equation}
where we tagged the various contributions explicitly. The amazing
fact is that the total universal beta function of $G_k$ is the same as in
the strictly two dimensional case (\ref{CG_6.3}), but now
the contributions are the following:
\begin{equation}
c_{g}=\underbrace{c_{\Phi}}_{\textrm{matter}}\:\underbrace{-10}_{\textrm{spin 2}}\:\underbrace{+7-6\xi}_{\textrm{spin 0}}\:\underbrace{-16}_{\textrm{ghost}}=\left\{ \begin{array}{c}
c_{\Phi}-19\qquad\xi=0\\
c_{\Phi}-25\qquad\xi=1
\end{array}\right.\,.\label{FG_17.1}
\end{equation}
Thus again $c_{g}=c_{\Phi}-25$ if $\xi=1$. Note that now the $-\frac{13}{3}$
contribution of the CG ghost is split up in a $-\frac{5}{3}$ from
``gravitons'' and a $-\frac{8}{3}$ from the FG ghost, this clearly
shows that the ghost contribution to $c_{g}$ alone has no universal
meaning. In both gauges, the trace behaves as a standard scalar only
in the exponential parametrization.
The same ghost contribution was found in the extrinsic approach, where one computes the RG flow in the theory of two dimensional surfaces embedded in $D$--dimensional Euclidean space and then takes the limit $D \to 0$  in the equivalent of Newton's constant beta function \cite{Codello:2011yf}. 

In terms of dimensionless variables, $\mbox{\ensuremath{\Lambda}}_{k}=k^{2}\tilde{\Lambda}_{k}$, and employing Litim's cutoff,
the beta functions (\ref{FG_16}) become:
\begin{eqnarray}
\partial_{t}\tilde{\Lambda}_{k} & = & -2\tilde{\Lambda}_{k}+\frac{1}{4\pi}\left\{ \frac{2}{1-(\xi-1)\tilde{\Lambda}_{k}}+\frac{1}{1-\frac{\xi}{\epsilon}\tilde{\Lambda}_{k}}-2\right\} \nonumber \\
\partial_{t}\left(-\frac{1}{G_{k}}\right) & = & -\frac{5}{3}\frac{1}{1-(\xi-1)\tilde{\Lambda}_{k}}+\frac{7-6\xi}{2}\frac{1}{1-\frac{\xi}{\epsilon}\tilde{\Lambda}_{k}}-\frac{8}{3}\,.\label{FG_18}
\end{eqnarray}
%
%
%
Note that only if the cosmological constant is zero we can take the limit
$d\rightarrow2$ when $\xi=1$. A more detail discussion of these beta functions can be found in the literature \cite{Codello:2008vh,Nink_2014}.

\subsubsection{Finite part of the effective action}

In this section we compute the finite parts of the effective action.
We will follow the methods of \cite{Codello_2010} which employ the non--local
heat kernel expansion, in particular we compute the $R^{2}$ terms.
As we have seen in the previous section, the Polyakov coupling does
not run, i.e. $\partial_{t}c_{k}=0$ (we need terms proportional to
beta functions to have a non--zero running). In this section we will
show that it is generated in the $k\rightarrow0$ limit (as explained
in \cite{Codello_2010}).

The non--local ansatz for the gauge invariant part of the effective
action to use one the lhs of the flow equation is:
\begin{equation}
\Gamma_{k}[g]=\int d^{2}x\sqrt{g}\left[a_{k}+b_{k}R+R\, c_{k}(\Delta)R\right]+O(R^{3})\,.\label{FP_1}
\end{equation}
In two dimensions the Ricci tensor is proportional to the Ricci scalar
$R_{\mu\nu}=\frac{1}{2}g_{\mu\nu}R$ so there is only one non--local
heat kernel structure function at the order curvature square \cite{Codello_Zanusso_2013} and this
is given by the following linear combination:
\begin{eqnarray}
f_{R^{2}}(x) & = & \textrm{tr}\,\mathbf{1}\, f_{R2d}(x)+\textrm{tr}\,\mathbf{1}\, f_{UR}(x)\left(\frac{d^{2}-3d+4}{d(d-1)}-\frac{d-2}{4d}\xi\right)\nonumber \\
 &  & +\textrm{tr}\,\mathbf{1}\, f_{U}(x)\left(\frac{d^{2}-3d+4}{d(d-1)}-\frac{d-2}{4d}\xi\right)^{2}+\left(1-\frac{4}{d}\right)(d+2)\, f_{\Omega}(x)\nonumber \\
 &  & +f_{R2d}(x)+f_{UR}(x)\frac{d-4+2\xi}{d}+f_{U}(x)\left(\frac{d-4+2\xi}{d}\right)^{2}\nonumber \\
 &  & +\delta_{\mu}^{\mu}\, f_{R2d}(x)-f_{UR}(x)\, U+\frac{1}{d}f_{U}(x)+\left(1-\frac{4}{d}\right)\, f_{\Omega}(x)\,.\label{FP_2}
\end{eqnarray}
The first two lines are the spin two contribution, the third line
is the spin zero contribution, while the last line is the spin one
contribution. When we write (\ref{FP_2}) in terms of the basic non--local
structure function and set $d=2$ we find:
\begin{eqnarray}
f_{R^{2}} & = & \frac{9f}{8x^{2}}+\frac{15f}{8x}+\frac{43f}{32}-\frac{9}{8x^{2}}-\frac{27}{16x}-\frac{5f}{4}\xi+\frac{1}{2x}\xi-\frac{f}{2x}\xi+\frac{f}{2}\xi^{2}\nonumber \\
 &  & +\frac{3f}{4x^{2}}+\frac{5f}{4x}+\frac{9f}{16}-\frac{3}{4x^{2}}-\frac{9}{8x}\,,\label{FP_3}
\end{eqnarray}
where not the first line is the gravitational contribution and the
second line the ghost contribution. The flow equation for $c_{k}(x)$
can be written as:
\begin{equation}
\partial_{t}c_{k}(x)=\frac{1}{8\pi k^{2}}g\left(\frac{x}{k^{2}}\right)\,.\label{FP_4}
\end{equation}
Note the overall power of $k^{-2}$ in (\ref{FP_4}). If we employ
Litim's cutoff shape function then we find ($u=x/k^{2}$):
\begin{eqnarray}
g(u) & = & -\frac{1}{8u^{2}}\left\{ \left[12+(27-8\xi)u\right]\sqrt{1-\frac{4}{u}}-(43-40\xi+16\xi^{2})u\sqrt{\frac{u}{u-4}}\right\} \theta(u-4)\nonumber \\
 &  & +\frac{1}{2u^{2}}\left[(4+9u)\sqrt{1-\frac{4}{u}}-9u\sqrt{\frac{u}{u-4}}\right]\theta(u-4)\,,\label{FP_5}
\end{eqnarray}
where the first line is the graviton contribution and the second line
is the ghost contribution. An expansion around $u=\infty$, 
\begin{equation}
g(u)=\frac{2\xi^{2}-4\xi+2}{u}+\frac{4\xi^{2}-12\xi+16}{u^{2}}+O\left(\frac{1}{u^{3}}\right)\,,\label{FP_6}
\end{equation}
shows the important point that in the exponential parametrization
the coefficient of the leading term is zero, i.e. $2\xi^{2}-4\xi+2=0$
if $\xi=1$. This fact was also noticed in the original covariant
perturbation theory literature in the case of a scalar field \cite{Vilkovisky1992za}.
We will see in a moment that this behaviour makes the integral of
$g(u)$ finite in the $k\rightarrow0$ limit. This cancellation can
be seen directly in (\ref{FP_5}) since the coefficients of the $u$
times square roots terms agree when $\xi=1$.

Integrating the flow from the UV scale $\Lambda$ to the IR scale
$k$ and shifting to the variable $u=x/k^{2}$ gives:
\begin{equation}
c_{k}(x)=c_{\Lambda}(x)-\frac{1}{16\pi x}\int_{x/\Lambda^{2}}^{x/k^{2}}du\, g(u)\,.\label{FP_7}
\end{equation}
The integral in (\ref{FP_7}) is finite for $\Lambda\rightarrow\infty$,
i.e. there are no UV divergences; so we can take the UV cutoff to
infinity. This is related to the theta functions in (\ref{FP_5})
which imply that we have to compute the integral between $4$ and
$x/k^{2}$, explicitly showing that the high energy part does not
contribute. The integral in (\ref{FP_7}) can be performed analytically,
but the result is not very revealing and in any case scheme dependent.
Instead we report the small $k$ expansion:
\begin{eqnarray}
c_{k}(x)-c_{\infty}(x) & = & -\frac{1}{16\pi x}\left[\frac{13}{2}-2\xi+2(1-\xi)^{2}\log\frac{x}{k^{2}}-4\left(\xi^{2}-3\xi+4\right)\frac{k^{2}}{x}\right.\nonumber \\
 &  & \left.\qquad\qquad-\frac{26}{3}+16\frac{k^{2}}{x}+O\left(\frac{k^{4}}{x^{2}}\right)\right]\,,\label{FP_7.1}
\end{eqnarray}
where again we separated the gravitational from the ghost contributions.
This expression shows clearly that the limit $k\rightarrow0$ is obstructed
by the diverging logarithm term if $\xi\neq1$, i.e. there is an IR
divergence if we use the standard parametrization. In the exponential
parametrization, instead, we can safely take the IR limit to find:
\begin{equation}
c_{0}(x)=c_{\infty}(x)+\frac{25}{96\pi x}\equiv-\frac{c_{\infty}-25}{96\pi x}\,,\label{FP_8}
\end{equation}
thus $c_{0}=c_{\infty}-25$. This checks explicitly our assumptions
that the gravitational effective action has the same form as the Polyakov
effective action for matter fields, but with the proper coefficient $c_g=c_\Phi-25$ (with matter included as in \cite{Codello_2010}).

\section{Conclusions}

In this paper we have explored the quantum properties of two dimensional
quantum gravity by putting together two complementary approaches:
scaling arguments and the renormalization group (RG) analysis.
In both cases we pursued a fully covariant formulation to prepare the ground for a future study of four dimensional quantum gravity along the same lines.

The full quantum properties of a theory are only accessible when we
have a well defined quantization procedure. In the path integral approach,
which is the most useful one to set up the RG analysis, this translates into
correctly identifying the measure of integration. We found that by
using the prescription given in \cite{Codello_D'Odorico_Pagani_2014}, standard results
in two dimensional quantum gravity can be reproduced in a simple and
clear way.
Once the quantization is understood, a shortcut to study the UV properties
of gravity is to consider the partition function at a fixed volume,
which is an area in $d=2$, and study how this scales when we rescale
the area. The path integral will generate a nontrivial quantum scaling
on top of the classical one, which determines the quantum properties
of the theory. The standard partition function then can be recovered
from this by a Laplace transform, and it will pick up these contributions.
The advantage of this construction is that once this ``reduced''
partition function is well defined, exact scaling relations can be
derived from it which translate into exact properties of the full
partition function. In this way one is able to find out how gravity
modifies the spectrum of scaling dimensions of matter, and essentially
solve quantum gravity in two dimensions.

Scaling relations are a natural hint for RG arguments. A paradigmatic
example are the ones one finds in a statistical physics context. First
observed in purely phenomenological terms, they relate the various
critical exponents of a statistical system by assuming that the free
energy (read partition function) has some definite scaling form, in
terms of one dimensionful quantity, say the temperature, and a function
of dimensionless ratios. The exponent of the temperature is one of
the critical exponents. Other quantities derived from the free energy
can be put in similar scaling form, and by comparing these different
quantities scaling relations between different exponents are found.
The RG gives an intuitive reason for this. Since at the critical point
correlation lengths diverge, the system reaches a scale invariant
phase, which is associated to a fixed point of the RG flow. At the fixed
point all dimensionless quantities approach a finite value (including
zero), so physical observables acquire a definite scaling with respect
to a dimensionful scale. Moreover, by considering the linearized RG
flow in the neighbourhood of the fixed point, it is actually possible
to calculate the value of the critical exponents, whereas scaling
arguments alone are not sufficient to determine them.

This is conceptually the same approach that was followed here. To calculate
the gravitational scaling exponents, we used the functional RG.
The scaling exponents can then be easily found as the wave
function renormalization of composite operators.
%
Still, up to this point there is one information missing, which is
the value of the gravitational central charge. Here we found that
to reproduce the correct value, consistent with what one finds from
Liouville theory, one needs to use the exponential parametrization
for the metric fluctuations. We performed the calculation in $d=2$
and $d=2+\epsilon$. In the last case one also looks at the finite part
of the effective action, and recovers the same result. It is crucial
there to use the exponential parametrization in order to be able to
recover the $k\to0$ limit.
The result also agrees with the relation found in \cite{Codello_D'Odorico_Pagani_2014}
between the central charge and the beta function of Newton's constant,
which was briefly rederived, in a slightly different way, in the first
part of the paper. This relation allows us to use the known form of
the beta function to compute a universal quantity like the c--function.

Probably the most interesting aspect of this work was that Liouville
theory, which is peculiar to two dimensions, was never really needed.
All we needed were scaling arguments and RG calculations. As we said, the whole
approach seems to work in covariant form. Thus there is the hope that
the same analysis can be carried through in $d=4$. We will investigate
this in a future paper.

\paragraph*{Acknowledgements.} The Authors would like to thank R. Percacci for carefully reading the first draft of this work and for many interesting suggestions and comments.

\appendix

\section{Variations}

In this Appendix we collect the basic variations, and their derivations, which are needed in the main
text.

\subsection{Variations of $\sqrt{g}$ and $\sqrt{g}R$}

We start considering the variations of the two invariants
\begin{equation}
I_{0}[g]=\int\sqrt{g}\qquad\qquad I_{1}[g]=\int\sqrt{g}R\,.\label{X_1}
\end{equation}
To compute the Hessians entering the flow equation we need the second
variations of the invariants (\ref{X_1}).These read, taken for example
from \cite{Codello:2008vh} or computed using \texttt{xTensor}, are:
\begin{equation}
\delta^{2}I_{0}[g]=\int d^{d}x\sqrt{g}\left(\frac{1}{4}h^{2}-\frac{1}{2}h^{\alpha\beta}h_{\alpha\beta}+\frac{1}{2}H\right)\,,\label{X_2}
\end{equation}
where we remember that $H_{\mu\nu}=\delta^{2}g_{\mu\nu}=\xi h_{\mu\lambda}h_{\mu}^{\lambda}$,
and:
\begin{eqnarray}
\delta^{2}I_{1}[g] & = & \int d^{d}x\sqrt{g}\left[-\frac{1}{2}h^{\mu\nu}\Delta h_{\mu\nu}+\frac{1}{2}h\Delta h-h^{\mu\nu}\nabla_{\nu}\nabla_{\alpha}h_{\mu}^{\alpha}+h\nabla^{\mu}\nabla^{\nu}h_{\mu\nu}\right.\nonumber \\
 &  & +h^{\mu\nu}h_{\mu}^{\alpha}R_{\nu\alpha}+h^{\mu\nu}h^{\alpha\beta}R_{\alpha\mu\beta\nu}-hR^{\mu\nu}h_{\mu\nu}+\left(\frac{1}{4}h^{2}-\frac{1}{2}h^{\alpha\beta}h_{\alpha\beta}\right)R\nonumber \\
 &  & \left.-H_{\mu\nu}R^{\mu\nu}+\frac{1}{2}HR+\nabla_{\mu}\nabla_{\nu}H^{\mu\nu}+\Delta H\right]\,.\label{X_3}
\end{eqnarray}
We note that the last two terms are total derivatives, thus the use
of the exponential parametrization will not change the kinetic terms
of the standard parametrization and so the difference will be in the
curvature terms.

It is useful now to perform the trace--traceless decomposition,
\begin{equation}
h_{\mu\nu}=\hat{h}_{\mu\nu}+\frac{1}{d}h\bar{g}_{\mu\nu}\qquad\qquad\bar{g}^{\mu\nu}\hat{h}_{\mu\nu}=0\,,\label{X_4}
\end{equation}
so that:
\begin{eqnarray}
h^{\mu\nu}h_{\mu\nu} & = & \hat{h}^{\mu\nu}\hat{h}_{\mu\nu}+\frac{1}{d}h^{2}\nonumber \\
h^{\mu\nu}\Delta h_{\mu\nu} & = & \hat{h}^{\mu\nu}\Delta\hat{h}_{\mu\nu}+\frac{1}{d}h\Delta h\nonumber \\
h\nabla^{\mu}\nabla^{\nu}h_{\mu\nu} & = & h\nabla^{\mu}\nabla^{\nu}\hat{h}_{\mu\nu}-\frac{1}{d}h\Delta h\nonumber \\
h^{\mu\nu}\nabla_{\nu}\nabla_{\alpha}h_{\mu}^{\alpha} & = & \hat{h}^{\mu\nu}\nabla_{\nu}\nabla_{\alpha}\hat{h}_{\mu}^{\alpha}+\frac{2}{d}h\nabla^{\mu}\nabla^{\nu}\hat{h}_{\mu\nu}-\frac{1}{d^{2}}h\Delta h\,,\label{X_5}
\end{eqnarray}
where an integration by parts is implicit in the last relation. Using
these relations in (\ref{X_2}) gives:
\begin{equation}
\delta^{2}I_{0}[g]=\int d^{d}x\sqrt{g}\left(\frac{1}{4}h^{2}+\frac{\xi-1}{2}h^{\alpha\beta}h_{\alpha\beta}\right)=\int d^{d}x\sqrt{g}\left(\frac{d-2+2\xi}{4d}h^{2}+\frac{\xi-1}{2}\hat{h}^{\alpha\beta}\hat{h}_{\alpha\beta}\right)\,.\label{X_6}
\end{equation}
Thus in two dimensions the second variation of $I_{0}[g]$ is purely
traceless in the standard parametrization $\xi=0$, while it is purely
trace in the exponential parametrization $\xi=1$. Note that in the
standard parametrization there is a dangerous $d-2$ pole term that
must be properly treated in the $d\rightarrow2$ limit.

Using (\ref{X_5}) in (minus) the derivative terms of (\ref{X_3})
gives:
\[
\frac{1}{2}h^{\mu\nu}\Delta h_{\mu\nu}-\frac{1}{2}h\Delta h+h^{\mu\nu}\nabla_{\nu}\nabla_{\alpha}h_{\mu}^{\alpha}-h\nabla^{\mu}\nabla^{\nu}h_{\mu\nu}=\qquad
\]
\begin{equation}
\qquad\qquad\frac{1}{2}\hat{h}^{\mu\nu}\Delta\hat{h}_{\mu\nu}-\frac{(d-2)(d-1)}{2d^{2}}h\Delta h+\hat{h}^{\mu\nu}\nabla_{\nu}\nabla_{\alpha}\hat{h}_{\mu}^{\alpha}-\frac{d-2}{d}h\nabla^{\mu}\nabla^{\nu}\hat{h}_{\mu\nu}\,.\label{X_7}
\end{equation}
We will now choose the background metric to be maximally symmetry
in order to simplify the curvature terms of (\ref{X_3}); in two dimensions
this is no restriction at all, while in $d\geq2$ we loose no generality
since we are interested in expansions of the effective action up to
linear order in the curvature. The Riemann and Ricci tensors are then
proportional to the Ricci scalar:
\begin{equation}
R_{\mu\nu}=\frac{R}{d}g_{\mu\nu}\qquad\qquad R_{\mu\nu\rho\sigma}=\frac{R}{d(d-1)}\left(g_{\mu\rho}g_{\nu\sigma}-g_{\mu\sigma}g_{\nu\rho}\right)\,.\label{X_8}
\end{equation}
Inserting these relations and performing the trace--traceless decomposition
of (minus) the curvature terms of (\ref{X_3}) gives:
\[
-h^{\mu\nu}h_{\mu}^{\alpha}R_{\nu\alpha}-h^{\mu\nu}h^{\rho\alpha}R_{\rho\nu\alpha\mu}+hh_{\mu\nu}R^{\mu\nu}-\left(\frac{1}{4}h^{2}-\frac{1}{2}h^{\alpha\beta}h_{\alpha\beta}\right)R+H_{\mu\nu}R^{\mu\nu}-\frac{1}{2}HR\qquad
\]
\begin{equation}
\qquad=\frac{1}{2}\hat{h}^{\mu\nu}\hat{h}_{\mu\nu}\left(\frac{d^{2}-3d+4}{d(d-1)}-\frac{d-2}{d}\xi\right)R+h^{2}\left(\frac{d^{2}-3d+4}{2d^{2}(d-1)}-\frac{d-2}{4d^{2}}\xi-\frac{d^{2}-5d+8}{4d(d-1)}\right)R\,.\label{X_9}
\end{equation}
Finally we are led to:
\[
\delta^{2}I_{1}[g]=-\int d^{d}x\sqrt{g}\left\{ \frac{1}{2}\hat{h}^{\mu\nu}\left(\Delta+\frac{d^{2}-3d+4}{d(d-1)}R-\frac{d-2}{d}\xi R\right)\hat{h}_{\mu\nu}\right.\qquad\qquad
\]
\begin{equation}
\left.\qquad+\hat{h}^{\mu\nu}\nabla_{\nu}\nabla_{\alpha}\hat{h}_{\mu}^{\alpha}-\frac{d-2}{d}h\nabla^{\mu}\nabla^{\nu}\hat{h}_{\mu\nu}-\frac{d-2}{2d^{2}}h\left[(d-1)\Delta+\frac{d-4+2\xi}{2}R\right]h\right\} \,.\label{X_10}
\end{equation}
This relation is interesting for several reasons; first it shows that
the trace part vanishes in two dimension making the Hessian non--invertible;
second it shows that $\xi$--terms make no difference when again $d=2$.
These are both signs of the topological nature of the operator in
this dimension and imply that in CG the contribution of the invariant
$I_{1}[g]$ is purely gauge and will not contribute when we will enforce
the gauge strictly (as expected). In FG, as for the other invariant,
the fact that the inverse of the trace part is singular when $d\rightarrow2$
calls for a careful definition of the regulator.

The background FG gauge fixing action (\ref{PI_1.4}) is already quadratic
in $h_{\mu\nu}$, when expanded it reads:
\begin{eqnarray}
S_{gf}[h;g] & = & \frac{1}{2\alpha}\int d^{d}x\sqrt{g}\left(-h^{\mu\nu}\nabla_{\nu}\nabla_{\alpha}h_{\mu}^{\alpha}+h\nabla^{\mu}\nabla^{\nu}h_{\mu\nu}+\frac{1}{4}h\Delta h\right)\nonumber \\
 & = & \frac{1}{2\alpha}\int d^{d}x\sqrt{g}\left(-\hat{h}^{\mu\nu}\nabla_{\nu}\nabla_{\alpha}\hat{h}_{\mu}^{\alpha}+\frac{d-2}{d}h\nabla^{\mu}\nabla^{\nu}\hat{h}_{\mu\nu}+\frac{(d-2)^{2}}{4d^{2}}h\Delta h\right)\,.\label{X_11}
\end{eqnarray}
Combining now (\ref{X_10}) with (\ref{X_11}) gives:
\begin{eqnarray}
-\frac{1}{2}\delta^{2}I_{1}[g]+S_{gf}[h;g] & = & \frac{1}{2}\int d^{d}x\sqrt{g}\left[\frac{1}{2}\hat{h}^{\mu\nu}\Delta\hat{h}_{\mu\nu}-\frac{d-2}{2d^{2}}\left(d-1-\frac{d-2}{2\alpha}\right)h\Delta h\right.\nonumber \\
 &  & +\left(1-\frac{1}{\alpha}\right)\hat{h}^{\mu\nu}\nabla_{\nu}\nabla_{\alpha}\hat{h}_{\mu}^{\alpha}-\frac{d-2}{d}\left(1-\frac{1}{\alpha}\right)h\nabla^{\mu}\nabla^{\nu}\hat{h}_{\mu\nu}\nonumber \\
 &  & +\hat{h}^{\mu\nu}\hat{h}_{\mu\nu}\left(\frac{d^{2}-3d+4}{2d(d-1)}-\frac{d-2}{2d}\xi\right)R\nonumber \\
 &  & \left.-h^{2}\frac{(d-2)(d-4+2\xi)}{4d^{2}}R\right]\,,\label{X_12}
\end{eqnarray}
which shows that the choice $\alpha=1$ leads to the diagonalization
of the quadratic action:
\[
-\frac{1}{2}\delta^{2}I_{1}[g]+S_{gf}[h;g]=\frac{1}{2}\int d^{d}x\sqrt{g}\left\{ \frac{1}{2}\hat{h}^{\mu\nu}\left[\Delta+\left(\frac{d^{2}-3d+4}{d(d-1)}R-\frac{d-2}{d}\xi\right)R\right]\hat{h}_{\mu\nu}\right.
\]
\begin{equation}
\left.-\frac{d-2}{4d}h\left[\Delta+\frac{d-4+2\xi}{d}R\right]h\right\} \,.\label{X_13}
\end{equation}
We stress again that the Hessian of this quadratic action will not
be invertible in two dimension since the trace part vanishes.

\subsection{Variations of the Polyakov action}

We will now determine the Hessian of the Polyakov action (\ref{M_2}):
\begin{equation}
S_{P}[g]=-\frac{1}{96\pi}\int\sqrt{g}\, R\frac{1}{\Delta}R\,.\label{A_1}
\end{equation}
First we will need some basic variations:
\begin{eqnarray}
\delta g^{\mu\nu} & = & -h^{\mu\nu}\nonumber \\
\delta\sqrt{g} & = & \frac{1}{2}\sqrt{g}h\nonumber \\
\delta\Gamma_{\mu\nu}^{\alpha} & = & \frac{1}{2}\left(\nabla_{\mu}h_{\;\nu}^{\alpha}+\nabla_{\nu}h_{\;\mu}^{\alpha}-\nabla^{\alpha}h_{\mu\nu}\right)\nonumber \\
g^{\mu\nu}\delta\Gamma_{\mu\nu}^{\alpha} & = & \nabla^{\mu}h_{\;\mu}^{\alpha}-\frac{1}{2}\nabla^{\alpha}h\nonumber \\
\delta R & = & \Delta h+\nabla^{\mu}\nabla^{\nu}h_{\mu\nu}-\frac{1}{2}hR\,,\label{A_2}
\end{eqnarray}
where we used $R_{\mu\nu}=\frac{1}{2}g_{\mu\nu}R$ in the last line,
relation valid only in two dimensions. We start with
\begin{equation}
\delta\left(\sqrt{g}\, R\frac{1}{\Delta}R\right)=\frac{1}{2}h\, R\frac{1}{\Delta}R+2\,\delta R\frac{1}{\Delta}R+R\,\delta\frac{1}{\Delta}R\,,\label{A_3}
\end{equation}
where we integrated by parts one of the terms with the variation of
the Ricci scalar. From $\frac{1}{\Delta}\Delta=1$ we find
\begin{equation}
\delta\frac{1}{\Delta}=-\frac{1}{\Delta}\delta\Delta\frac{1}{\Delta}\,,\label{A_4}
\end{equation}
where the variation of the Laplacian acting on a scalar, using variations
form (\ref{A_2}), is:
\begin{eqnarray}
\delta\Delta\phi & = & -\delta\left(g^{\mu\nu}\nabla_{\nu}\nabla_{\mu}\phi\right)\nonumber \\
 & = & -\delta g^{\mu\nu}\nabla_{\nu}\nabla_{\mu}\phi-g^{\mu\nu}\delta\nabla_{\mu}\partial_{\nu}\phi\nonumber \\
 & = & h^{\mu\nu}\nabla_{\nu}\nabla_{\mu}\phi+g^{\mu\nu}\delta\Gamma_{\mu\nu}^{\alpha}\nabla_{\alpha}\phi\nonumber \\
 & = & h^{\mu\nu}\nabla_{\mu}\nabla_{\nu}\phi+\nabla^{\mu}h_{\;\mu}^{\nu}\nabla_{\nu}\phi-\frac{1}{2}\nabla^{\alpha}h\nabla_{\alpha}\phi\,.\label{A_5}
\end{eqnarray}
Inserting (\ref{A_5}) in (\ref{A_4}) and then (\ref{A_4}) in (\ref{A_3})
gives:
\begin{eqnarray}
\frac{1}{\sqrt{g}}\delta\left(\sqrt{g}\, R\frac{1}{\Delta}R\right) & = & -\frac{1}{2}h\, R\frac{1}{\Delta}R+2hR+2h_{\mu\nu}\nabla^{\mu}\nabla^{\nu}\frac{1}{\Delta}R+h^{\mu\nu}\left(\nabla_{\mu}\frac{1}{\Delta}R\right)\left(\nabla_{\nu}\frac{1}{\Delta}R\right)+\nonumber \\
 &  & -\frac{1}{2}h\left(\nabla^{\alpha}\frac{1}{\Delta}R\right)\left(\nabla_{\alpha}\frac{1}{\Delta}R\right)+\frac{1}{2}hR\frac{1}{\Delta}R\,.\label{A_6}
\end{eqnarray}
Equation (\ref{A_6}) gives us directly the the first variation of
the Polyakov action: 
\begin{equation}
-96\pi\delta S_{P}[g]=\int d^{2}x\sqrt{g}\left\{ hR+\hat{h}_{\mu\nu}\left[2\nabla^{\mu}\nabla^{\nu}\frac{1}{\Delta}R+\left(\nabla^{\mu}\frac{1}{\Delta}R\right)\left(\nabla^{\nu}\frac{1}{\Delta}R\right)\right]\right\} \,,\label{A_7}
\end{equation}
where we have separated the trace part from trace free part. Defining
the tensor
\begin{equation}
t^{\mu\nu}=2\nabla^{\mu}\nabla^{\nu}\frac{1}{\Delta}R+\left(\nabla^{\mu}\frac{1}{\Delta}R\right)\left(\nabla^{\nu}\frac{1}{\Delta}R\right)\,,\label{A_8.1}
\end{equation}
allows us to write (\ref{A_7}) in the following simple way:
\begin{equation}
\delta S_{P}[g]=-\frac{1}{96\pi}\int d^{2}x\sqrt{g}\left\{ hR+\hat{h}_{\mu\nu}t^{\mu\nu}\right\} \,.\label{A_8}
\end{equation}
From this relation we can compute the energy--momentum tensor,
\begin{equation}
\left\langle T^{\mu\nu}\right\rangle =\frac{2}{\sqrt{g}}\frac{\delta S_{P}[g]}{\delta g_{\mu\nu}}=-\frac{1}{48\pi}\left\{ g^{\mu\nu}R+t^{\mu\nu}-\frac{1}{2}g^{\mu\nu}t\right\} \,,\label{A_8.01}
\end{equation}
where $t\equiv t_{\;\alpha}^{\alpha}=-2R+\left(\nabla^{\mu}\frac{1}{\Delta}R\right)\left(\nabla_{\mu}\frac{1}{\Delta}R\right)$,
which shows directly the anomalous trace
\begin{equation}
g_{\mu\nu}\left\langle T^{\mu\nu}\right\rangle =-\frac{R}{24\pi}\,.\label{A_9}
\end{equation}
Now note that we could had written (\ref{A_3}) as:
\[
\frac{1}{\sqrt{g}}\delta\left(\sqrt{g}\, R\frac{1}{\Delta}R\right)=\frac{1}{2}h\, R\frac{1}{\Delta}R+\delta R\frac{1}{\Delta}R+R\,\delta\left(\frac{1}{\Delta}R\right)\,;
\]
since
\[
\frac{1}{2}h\, R\frac{1}{\Delta}R+\delta R\frac{1}{\Delta}R=\frac{1}{2}hR+\hat{h}_{\mu\nu}\nabla^{\mu}\nabla^{\nu}\frac{1}{\Delta}R\,,
\]
it must be that:
\begin{equation}
\delta\left(\frac{1}{\Delta}R\right)=\frac{1}{2}h+\frac{1}{R}\hat{h}_{\mu\nu}\left[\nabla^{\mu}\nabla^{\nu}\frac{1}{\Delta}R+\left(\nabla^{\mu}\frac{1}{\Delta}R\right)\left(\nabla^{\nu}\frac{1}{\Delta}R\right)\right]\,.\label{A_10}
\end{equation}
thus $\sigma(g)=\frac{1}{2\Delta}R$ in strict CG ($\hat{h}_{\mu\nu}=0$)
is such that $\delta\sigma=\frac{1}{4}h$. Note also that in strict
CG the first variation (\ref{A_8}) becomes:
\begin{equation}
\delta S_{P}[g]=-\frac{1}{96\pi}\int d^{2}x\sqrt{g}hR\,.\label{A_11}
\end{equation}
We can now perform the second variation of (\ref{A_11}) directly
(we leave to the next Section the second variation in general gauge).
Using the basic variations (\ref{A_2}), the traceless--trace decomposition
relations (\ref{X_5}) and $\delta(\sqrt{g}\hat{h}_{\mu\nu})=\frac{1}{2}\hat{h}^{\mu\nu}\hat{h}_{\mu\nu}g_{\mu\nu}$
we find:
\begin{equation}
\delta^{2}S_{P}[g]=-\frac{1}{96\pi}\int d^{2}x\,\sqrt{g}\left\{ \frac{1}{2}h\Delta h-\frac{1}{2}h^{2}R+h\nabla^{\mu}\nabla^{\nu}\hat{h}_{\mu\nu}+\hat{h}^{\mu\nu}\hat{h}_{\mu\nu}\left(\frac{1}{2}t-R\right)+\hat{h}_{\mu\nu}\delta t^{\mu\nu}\right\} \,.\label{A_12}
\end{equation}
We have not written explicitly the variation $\delta t^{\mu\nu}$
since it is not very illuminating and in any case we will not need
the explicit expression for it; we use the fact that $\delta t^{\mu\nu}$
has both traceless and trace parts to write (\ref{A_12}) as:
\begin{equation}
\delta^{2}S_{P}[g]=-\frac{1}{96\pi}\frac{1}{2}\int d^{2}x\,\sqrt{g}\left\{ h(\Delta-R)h+2hA^{\mu\nu}\hat{h}_{\mu\nu}+\hat{h}_{\mu\nu}B_{\alpha\beta}^{\mu\nu}\hat{h}^{\alpha\beta}\right\} \,,\label{A_13}
\end{equation}
where $A^{\mu\nu}$ and $B_{\alpha\beta}^{\mu\nu}$ can be read off
from (\ref{A_12}) and the knowledge of $\delta t^{\mu\nu}$.

We now consider the variation leading to terms proportional to $H_{\mu\nu}\equiv\delta^{2}g_{\mu\nu}=\xi h_{\mu\lambda}h_{\nu}^{\lambda}$:
\begin{eqnarray*}
\delta\int d^{2}x\sqrt{g}R\frac{1}{\Delta}R & = & 2\int d^{2}x\sqrt{g}R\frac{1}{\Delta}\delta R+O(R^{2})\\
 & = & 2\int d^{2}x\sqrt{g}R\frac{1}{\Delta}\left(\Delta H+\nabla^{\mu}\nabla^{\nu}H_{\mu\nu}-\frac{1}{2}HR\right)+O(R^{2})\\
 & = & 2\int d^{2}x\sqrt{g}\left(RH+R\frac{1}{\Delta}\nabla^{\mu}\nabla^{\nu}H_{\mu\nu}\right)+O(R^{2})\\
 & = & 2\int d^{2}x\sqrt{g}H_{\mu\nu}\left(Rg^{\mu\nu}+\nabla^{\mu}\nabla^{\nu}\frac{1}{\Delta}R\right)+O(R^{2})\,.
\end{eqnarray*}
The trace part is:
\begin{eqnarray}
2\int d^{2}x\sqrt{g}H_{\mu\nu}\left(Rg^{\mu\nu}+\nabla^{\mu}\nabla^{\nu}\frac{1}{\Delta}R\right)
 & = & \xi\int d^{2}x\sqrt{g}\frac{1}{2}h^{2}R\,.\label{A_14}
\end{eqnarray}
Combining this with (\ref{A_13}) finally gives:
\begin{equation}
\delta^{2}S_{P}[g]=-\frac{1}{96\pi}\frac{1}{2}\int d^{2}x\,\sqrt{g}\left\{ h\left[\Delta+(\xi-1)R\right]h+2hA^{\mu\nu}\hat{h}_{\mu\nu}+\hat{h}_{\mu\nu}B_{\alpha\beta}^{\mu\nu}\hat{h}^{\alpha\beta}\right\} \,,\label{A_15}
\end{equation}
with new $A^{\mu\nu}$ and $B_{\alpha\beta}^{\mu\nu}$. This
completes the collection of variations that we need.

As a final check of (\ref{A_11}) and (\ref{A_15}), we correctly
find that in strict CG and for $\xi=1$ the following relation between
variations of the Polyakov and Liouville actions is fulfilled:
\begin{equation}
\delta S_{P}[g]+\frac{1}{2}\delta^{2}S_{P}[g]=-\frac{1}{24\pi}\int d^{2}x\sqrt{g}\left\{ \frac{h}{4}R+\frac{h}{4}\Delta\frac{h}{4}\right\} =-\frac{1}{24\pi}S_{L}\left[\frac{h}{4};g\right]\,.\label{A_16}
\end{equation}
This relation strongly supports the use of the exponential parametrization.

Finally, for completeness we report the variation of $\delta t^{\mu\nu}$
for which we will need the following relations:
\begin{eqnarray}
\delta\left(\nabla_{\mu}\nabla_{\nu}\frac{1}{\Delta}R\right) & = & \delta\left(\nabla_{\mu}\partial_{\nu}\frac{1}{\Delta}R\right)\nonumber \\
 & = & \delta\left(\partial_{\mu}\partial_{\nu}\frac{1}{\Delta}R+\Gamma_{\mu\nu}^{\lambda}\partial_{\lambda}\frac{1}{\Delta}R\right)\nonumber \\
 & = & \partial_{\mu}\partial_{\nu}\delta\left(\frac{1}{\Delta}R\right)+\delta\Gamma_{\mu\nu}^{\lambda}\partial_{\lambda}\left(\frac{1}{\Delta}R\right)+\Gamma_{\mu\nu}^{\lambda}\partial_{\lambda}\delta\left(\frac{1}{\Delta}R\right)\nonumber \\
 & = & \nabla_{\mu}\nabla_{\nu}\delta\left(\frac{1}{\Delta}R\right)+\frac{1}{2}\left(\nabla_{\mu}h_{\;\nu}^{\lambda}+\nabla_{\nu}h_{\;\mu}^{\lambda}-\nabla^{\lambda}h_{\mu\nu}\right)\nabla_{\lambda}\left(\frac{1}{\Delta}R\right)\label{A_17}
\end{eqnarray}
and
\begin{equation}
\delta\left(\nabla_{\mu}\frac{1}{\Delta}R\right)=\partial_{\mu}\delta\left(\frac{1}{\Delta}R\right)=\nabla_{\mu}\delta\left(\frac{1}{\Delta}R\right)\,.\label{A_18}
\end{equation}
Using these relations and (\ref{A_10}) in the definition (\ref{A_8.1})
gives:
\begin{eqnarray}
\hat{h}_{\mu\nu}\delta t^{\mu\nu} & = & \hat{h}^{\mu\nu}\delta\left(\nabla_{\mu}\nabla_{\nu}\frac{1}{\Delta}R\right)+\hat{h}_{\mu\nu}\left(\nabla^{\mu}\frac{1}{\Delta}R\right)\nabla^{\nu}\delta\left(\frac{1}{\Delta}R\right)\nonumber \\
 & = & \frac{1}{2}\hat{h}^{\mu\nu}\nabla_{\mu}\nabla_{\nu}h+\hat{h}^{\mu\nu}\nabla_{\mu}\nabla_{\nu}\left\{ \frac{1}{R}\hat{h}_{\mu\nu}\left[\nabla^{\mu}\nabla^{\nu}\frac{1}{\Delta}R+\left(\nabla^{\mu}\frac{1}{\Delta}R\right)\left(\nabla^{\nu}\frac{1}{\Delta}R\right)\right]\right\} \nonumber \\
 &  & +\hat{h}^{\mu\nu}\left(\nabla_{\mu}\hat{h}_{\;\nu}^{\lambda}-\frac{1}{2}\nabla_{\mu}\delta_{\;\nu}^{\lambda}h\right)\nabla_{\lambda}\left(\frac{1}{\Delta}R\right)+\frac{1}{2}\hat{h}_{\mu\nu}\left(\nabla^{\mu}\frac{1}{\Delta}R\right)\nabla^{\nu}h\nonumber \\
 &  & +\hat{h}_{\mu\nu}\left(\nabla^{\mu}\frac{1}{\Delta}R\right)\nabla^{\nu}\left\{ \frac{1}{R}\hat{h}_{\mu\nu}\left[\nabla^{\mu}\nabla^{\nu}\frac{1}{\Delta}R+\left(\nabla^{\mu}\frac{1}{\Delta}R\right)\left(\nabla^{\nu}\frac{1}{\Delta}R\right)\right]\right\} \,,\label{A_19}
\end{eqnarray}
from which one can extract the explicit for of the tensors $A^{\mu\nu}$
and $B_{\alpha\beta}^{\mu\nu}$ appearing in (\ref{A_15}), but their
form is not very illuminating.


\begin{thebibliography}{10}




\bibitem{Berges:2000ew}
  J.~Berges, N.~Tetradis and C.~Wetterich,
  Phys.\ Rept.\  {\bf 363} (2002) 223
  
\bibitem{Reuter:1996cp}
  M.~Reuter,
  Phys.\ Rev.\ D {\bf 57} (1998) 971
  [hep-th/9605030].

\bibitem{KPZ}V. G. Knizhnik, A. M. Polyakov and A. B. Zamolodchikov,
Mod. Phys. Lett. A 3 (1988) 819.


\bibitem{Ambjorn:1997di}
  J.~Ambjorn, B.~Durhuus and T.~Jonsson,
  Cambridge, UK: Univ. Pr., 1997. (Cambridge Monographs in Mathematical Physics). 363 p


\bibitem{Codello_D'Odorico_Pagani_2014}A. Codello, G. D'Odorico and
C. Pagani, JHEP 1407 (2014) 040 {[}arXiv:1312.7097 {[}hep-th{]}{]}.

\bibitem{Distler_Kawai_1988}J. Distler and H. Kawai, Nucl. Phys.
B 321 (1989) 509.


\bibitem{David_1988}F. David, Mod. Phys. Lett. A 3 (1988) 1651.


%


  


\bibitem{Eichhorn:2013isa}
  A.~Eichhorn and T.~Koslowski,
  Phys.\ Rev.\ D {\bf 88} (2013) 084016
  [arXiv:1309.1690 [gr-qc]].
  
  A.~Eichhorn and T.~Koslowski,
  Phys.\ Rev.\ D {\bf 90} (2014) 10,  104039
  [arXiv:1408.4127 [gr-qc]].
  
  
\bibitem{Watabiki:1993fk}
  Y.~Watabiki,
  Prog.\ Theor.\ Phys.\ Suppl.\  {\bf 114} (1993) 1.
  
  \bibitem{Codello_Zanusso_2013}A. Codello and O. Zanusso, J. Math.
Phys. 54 (2013) 013513 {[}arXiv:1203.2034 {[}math-ph{]}{]}.

  
\bibitem{Reuter:2011ah}
  M.~Reuter and F.~Saueressig,
  JHEP {\bf 1112} (2011) 012
  [arXiv:1110.5224 [hep-th]].
  
  
\bibitem{Mottola:1995sj}
  E.~Mottola,
  J.\ Math.\ Phys.\  {\bf 36} (1995) 2470
  [hep-th/9502109].
  
  \bibitem{Kawai_Kitazawa_Ninomiya_1992}H. Kawai, Y. Kitazawa and M.
Ninomiya, Nucl. Phys. B 393 (1993) 280 {[}hep-th/9206081{]}.


\bibitem{Eichhorn:2013xr}
  A.~Eichhorn,
  Class.\ Quant.\ Grav.\  {\bf 30} (2013) 115016
  [arXiv:1301.0879 [gr-qc]].


\bibitem{Nink_2014}
A. Nink, arXiv:1410.7816 {[}hep-th{]}.

\bibitem{Codello:2008vh}
  A.~Codello, R.~Percacci and C.~Rahmede,
  Annals Phys.\  {\bf 324} (2009) 414
  [arXiv:0805.2909 [hep-th]].
  
   
\bibitem{Codello_2010}A. Codello, Annals Phys. 325 (2010) 1727 {[}arXiv:1004.2171
{[}hep-th{]}{]};
%
A. Satz, A. Codello and F.
D. Mazzitelli, Phys. Rev. D 82 (2010) 084011 {[}arXiv:1006.3808 {[}hep-th{]}{]};
A. Codello, New J. Phys. 14 (2012) 015009 {[}arXiv:1108.1908
{[}gr-qc{]}{]}.

\bibitem{Kawai_Ninomiya_1990}H. Kawai and M. Ninomiya, Nucl. Phys.
B 336 (1990) 115.

\bibitem{key-7}R. Gastmans, R. Kallosh and C. Truffin,
Nucl. Phys. B 133 (1978) 417; S. M. Christensen and M. J. Duff, Phys.
Lett. B 79 (1978) 213.
  
\bibitem{Codello:2013fpa}
  A.~Codello, G.~D'Odorico and C.~Pagani,
  Phys.\ Rev.\ D {\bf 89} (2014) 081701
  [arXiv:1304.4777 [gr-qc]].
  
\bibitem{Codello:2011yf}
  A.~Codello and O.~Zanusso,
  Phys.\ Rev.\ D {\bf 83} (2011) 125021
  [arXiv:1103.1089 [hep-th]].





\bibitem{Vilkovisky1992za}
  G.~A.~Vilkovisky,
  CERN-TH-6392-92.
  


\end{thebibliography}
\end{document}